\documentclass[11pt,a4paper]{article}
\pdfoutput=1

\usepackage{lineno}

\usepackage{jcappub}

\usepackage{listings}
\usepackage{multirow}
\usepackage{tikz}
\usepackage{epstopdf}
\usepackage{physics}
\usepackage{longtable}
\usepackage{setspace}
\usepackage{wasysym}
\usepackage{lscape}

\usepackage{times}

\usepackage[section]{placeins}

\usepackage{amssymb,amsmath,amsfonts}
\usepackage{graphicx,subfigure}
\usepackage{color}
\usepackage{rotating}
\usepackage{float}
\usepackage{epstopdf}
\usepackage{graphicx}
\usepackage{epsfig} 

\newcommand{\be}{\begin{equation}}
\newcommand{\ee}{\end{equation}}

\def \aap{Astron.~\&~Astrophys.}
\def \apjl{Astrophys.~J.~Lett.}

\def \apjs{ApJS}

\def\lesssim{\lower.5ex\hbox{$\; \buildrel < \over \sim \;$}}
\def\gtrsim{\lower.5ex\hbox{$\; \buildrel > \over \sim \;$}}
\newcommand{\lsim}{\,\rlap{\raise 0.35ex\hbox{$<$}}{\lower 0.7ex\hbox{$\sim$}}\,}
\newcommand{\gsim}{\,\rlap{\raise 0.35ex\hbox{$>$}}{\lower 0.7ex\hbox{$\sim$}}\,}

\newcommand{\bea}{\begin{eqnarray}}
\newcommand{\eea}{\end{eqnarray}}

\title{Consistency Between the Luminosity Function of Resolved Millisecond Pulsars and the Galactic Center Excess.
}

\author[a]{Harrison Ploeg}
\author[a]{Chris Gordon}
\author[b]{Roland Crocker} 
\author[c]{Oscar Macias}

\affiliation[a]{Department of Physics and Astronomy, Rutherford Building, University of Canterbury, Private Bag 4800, Christchurch 8140, New Zealand}
\affiliation[b]{Research School of Astronomy and Astrophysics, Australian National University, Canberra, Australia}
\affiliation[c]{Center for Neutrino Physics, Department of Physics, Virginia Tech, Blacksburg, VA 24061, USA}

\abstract{Fermi Large Area Telescope data reveal an excess of GeV gamma rays from the direction of the Galactic Center and bulge. 
Several explanations have been proposed for this excess including an unresolved population of millisecond pulsars (MSPs) and self-annihilating dark matter. 
It has been claimed that a key discriminant for or against the MSP explanation can be extracted from the properties of the luminosity function describing this source population. Specifically, is the 
luminosity function of the putative MSPs in the Galactic Center consistent 
with that characterizing the resolved MSPs in the Galactic disk? 
To investigate this we have used a Bayesian Markov Chain Monte Carlo to evaluate the posterior distribution of the parameters of the MSP 
luminosity function describing both  resolved MSPs and the Galactic Center excess.
At variance with some other claims, our analysis reveals that, within current uncertainties, both data sets 
can be well fit with the same luminosity function.
}

\keywords{Galactic Center, Gamma Rays, Millisecond Pulsars}

\begin{document}

\maketitle

\flushbottom

\section{Introduction}
An extended $\gamma$-ray source has been found \citep{Goodenough2009gk,hooperlinden2011,AbazajianKaplinghat2012,GordonMacias2013,MaciasGordon2014,Abazajian2014,Daylan:2014,Zho2015,
CaloreCholisWeniger2015,Ajello2016,Karwin2016,Dmitri2017}
 in the  {\it Fermi} Large Area Telescope ({\it Fermi}-LAT) data covering the central $\sim 10^{\circ}$  of the Milky Way.
This Galactic Center Excess (GCE) signal has a spectral peak at about $2$ GeV and reaches a maximum intensity at the Galactic Center (GC) from where it falls off radially like $\propto r^{-2.4}$.
Given its morphological and spectral characteristics, 
the GCE might constitute the indirect signature of the self-annihilation of   dark matter particles %
distributed in a Navarro-Frenk-White (NFW) like density profile. 

However, recent statistical studies~\citep{Lee_etal:2016,Bartes_etal:2016,Fermi-LAT:2017yoi} may
have uncovered, as dim clusters of photons, 
a population of unresolved, point-like sources in the GCE $\gamma$-ray signal.
These studies thus suggest,
contrary to the dark matter hypothesis, that the GCE
is actually attributable to many dim, unresolved point sources, presumably  of stellar origin
(though note it has also been argued that the photon clusters are merely due to variations in the gamma-ray 
flux associated
with the small scale structure of the diffuse Galactic emission  \citep{Horiuchi2016}).
A natural explanation of these $\gamma$-ray sources (were they real and of stellar origin) is that they are millisecond pulsars  (MSPs) \citep{Abazajian:2010zy,AbazajianKaplinghat2012,MaciasGordon2014,YuanZhang2014} 
and/or young pulsars
\citep{OLeary2015} 
which both have GeV-peaked gamma-ray spectra (also see \cite{Wang2005}).

This prompts the immediate question: what is the origin of this putative MSP and/or pulsar population?
The pulsar hypothesis requires relatively recent star formation given the $\lsim$ few Myr $\gamma$-ray lifetimes of ordinary pulsars.
Such star formation is absent from most of the bulge except in the $r \lsim 100$ pc nuclear region; a young pulsar explanation of the GCE thus requires that the bulge be populated with pulsars that
are launched out of the nucleus. It has been claimed that this can be achieved rather naturally  by the pulsar's natal kicks \citep{OLeary2015}.
On the other hand, MSPs can be generated (in a number of ways: see below) from old stellar populations.
Rather generically, two broad classes of bulge stellar population might be at play here: the bulge field stars or the bulge stars that were born in high redshift globular clusters that have been accumulated into the inner Galaxy by dynamical friction and disrupted by tidal forces over the lifetime of the Milky Way \citep{BrandtKocsis2015}.
Note that while the fraction of the bulge stellar population that derives from disrupted globular cluster is only at the $\sim$ few percent level \cite{Gnedin2014}, observationally, globular cluster environments are orders of magnitude more efficient per unit stellar mass at producing MSPs \cite{Camilo2005} than field stellar populations.

A number of arguments \cite{HooperLinden2016,HooperMohlabeng2016,Haggard:2017lyq} have been raised against MSP explanations for the GCE, the most important being:
\begin{enumerate}

\item It has been argued that, in the process of being spun up to millisecond (ms) periods by accretion from a binary companion, an MSP progenitor system will experience a low-mass X-ray binary (LMXB) phase and that, given there is a necessary connection between the LMXB and MSP phases, the same relative numbers of LMXB to MSPs should be seen in the GC and bulge as in other environments. 
However, the ratio of LMXBs to putative MSPs is much smaller for the GCE region than, for instance, globular cluster environments \cite{cholis2014,Haggard:2017lyq}.

\item It has been argued that the efficiency 
with which a given stellar mass of (putative) disrupted bulge globular clusters would have to be converted into MSPs 
is implausibly higher than the efficiency with which extant globular clusters generate $\gamma$-ray MSPs
\cite{HooperLinden2016}.

\item It has been argued that the luminosity distribution of an MSP population whose unresolved, lower-luminosity members might plausibly explain the apparently diffuse GCE signal is inconsistent with the luminosity distribution of more local (Galactic disk), resolved $\gamma$-ray MSPs \cite{HooperMohlabeng2016}. In particular, it has been claimed that {\it Fermi} should detect many more bright MSPs from the vicinity of the GC/bulge were an MSP population to explain the GCE, whereas it detects none.

\end{enumerate}

In this paper we will deal mostly with point 3. Before addressing this below, however,  
we take the opportunity here to explain that neither points 1 or 2 constitutes a watertight criticisms of the MSP scenario.


With respect to point 1, first note that different evolutionary pathways leading to the formation of MSPs have been proposed \cite{BhattacharyaHeuvel1991,IvanovaHeinkeRasio2008}, that, globally, the relative importance of these pathways is unclear, and that different pathways may have different relative importance in different stellar environments.
In particular, much of the discussion to date regarding the plausibility of an MSP scenario to explain the GCE signal has implicitly adopted the `recycling' scenario for MSP creation \cite{Alpar1982} where an old neutron star accretes material from a binary companion (which is either congenital or dynamically captured) and is spun up to ms periods.

However, other paths to the formation of MSPs exist and these remain entirely plausible from a theoretical perspective and not excluded by any data.
For instance, MSPs may be formed
via the accretion induced collapse (AIC) of a massive (close to $M_\mathrm{Chandra}$) O-Ne-Mg white dwarf accreting from a (typically low mass) companion in a tight binary \cite{Michel1987,GrindlayBailyn1988, BhattacharyaHeuvel1991,FerrarioWickramasinghe2007,Hurley2010,Tauris2013,Smedley2015}.
In such systems, conservation of angular momentum in the collapse from white dwarf to neutron star results directly in a $\sim$ms period remnant
and flux conservation naturally leads to a final neutron star field of $\sim 10^8$ G in felicitous agreement with that required by observations\footnote{Another nice property of AIC-derived MSPs is that they are not expected to receive anything like as hefty a natal kick as that delivered to neutron stars resulting from core-collapse supernovae \cite{Podsiadlowski2004}. This may help to explain how so many MSPs can accumulate in globular clusters \cite{Ivanova2008} -- and perhaps in the inner bulge too -- despite their shallow gravitational potentials.}.
No LMXB phases is thus necessary {\it before} the MSP emerges in this scenario.
Mass loss (to the gravitational binding energy of the neutron star) incurred by the remnant during AIC may lead 
to orbit widening and an interruption to accretion.
It is possible that accretion from the binary companion may be reestablished but it is also possible that 
winds from the nascent neutron star may sufficiently ablate the companion and/or the orbital separation increased to such an extent that accretion is not reestablished \cite{GrindlayBailyn1988,Smedley2015} so that an LMXB phase never occurs.

Moreover, another plausible formation channel for (individual) MSPs is merger induced collapse (MIC) of two white dwarfs. 
In this case there will certainly be no remaining binary companion after MIC and hence no LMXB phase \cite{Ivanova2008}.

In any case, even for AIC channels that lead to the formation of MSPs and subsequently undergo an LMXB phase, 
the relationship between the number of  MSPs and the number of extant LMXBs may be quite different in the AIC scenario from the recycling scenario given the duration of the typical LMXB phases for these pathways may also be different.
In summary, the uncertainty concerning which channel might dominate GC and bulge MSP creation means that the number of GC and bulge LMXBs cannot reliably be used as an indicator of the number in this region.

With respect to point 2, note that this argument only pertains in the case that the  bulge MSP population putatively responsible for the GCE does derive from globular clusters; as remarked above, it might instead derive from bulge field stars (and these might produce MSPs with a different efficency and through a different dominant channel than in the local disk).
Indeed, recently there have been indications that the GCE is not spherically symmetric \citep{YangAharonian2016} and 
that it is, in fact, spatially correlated with both the X-shaped stellar over-density in the Galactic bulge and the nuclear stellar bulge
\citep{Macias2016}.
This finding would seem to militate against both the young pulsar scenario
and the disrupted globular cluster scenarios because there does not seem to be any good reason that pulsars kicked out of the nucleus or MSPs delivered out of disrupted globular clusters should end up in such an X-shaped distribution\footnote{The stars that end up in the stellar X likely derive from a buckling instability induced in the early Galactic disk by the Galactic bar; see \cite{Macias2016} and refs.~therein.}.

Finally, with respect to point 3, note that it has been claimed \cite{HooperLinden2016} that the observed $\gamma$-ray luminosity distributions of the globular cluster and the field MSP populations are similar\footnote{This seems somewhat in tension with the finding
that  globular cluster MSPs seem to spin faster and have high magnetic fields than field MSPs in the disk \cite{Konar2010}. However, these effects may be at least partially attributable to observational biases.}.
Thus, were it true that disrupted globular clusters were the only plausible source of an MSP population that could supply the GCE and were point 3 above also true, this would seem to rule out the MSP scenario.
However, we will argue below that, contra point 3, given the size of current uncertainties, the luminosity distributions of disk and putative GCE MSPs are perfectly consistent with each other.
In addition, as we have already noted, this population need not necessarily derive from disrupted globular clusters (and, indeed, this scenario seems to be precluded by other considerations).

Given this, a putative MSP population that is responsible for the GCE presumably derives from the bulge field stars.
This is a stellar population quite different from that of the  local disk.
In particular, the vast majority of Galactic bulge (excluding the nucleus) stars  are $\gsim$ 8 Gyr old \cite{Nataf2016}  whereas the Galactic disk environment has experienced star formation up to the present day.
This difference may well mean that the stellar progenitors of the putative GCE region MSPs are systematically older than the progenitors of local MSPs (and, indeed, that the MSP population is systematically older and dimmer given that MSPs spin down and become less luminous over time).
There is then no strong reason to expect that the luminosity distributions of the local disk and the GCE region are the same because the stellar populations of these environments are quite different.

To this it may be objected that the stellar populations of MSP-hosting globular clusters are at least as old as the stars in the bulge.
However, globular clusters' typically much higher stellar densities may render the dynamic formation of binary systems
suitable for forming MSPs probable while a dynamic formation channel remains unlikely amongst the bulge field stars\footnote{Likewise, while the inner $\sim 200$ pc of the Galaxy does experience ongoing star formation, as mentioned, its stellar density and extreme interstellar medium parameters mean that it is a quite different environment to the local disk. Thus even though both regions experience ongoing star formation, they are also quite different environments.}.
In summary, the stellar populations and environments of the local disk, the Galactic bulge, globular clusters, and the Galactic nucleus are all quite distinct; there is no logical reason to expect that the populations of MSPs formed out of these different stellar populations/environments should be identical in character, particularly in regard to their current luminosity distribution.

Equally, given the stellar populations and stellar environments of the local disk and the bulge are quite distinct, the efficiency for the production of MSPs per unit stellar mass formed may be quite different (particularly if different MSP production channels are more or less in important in different circumstances).

\subsection{Plan of paper}

Having thus explained how the MSP scenario is by no means excluded by other considerations, in the remainder of this paper
we proceed to determine the luminosity function that simultaneously describes the population of  MSPs currently resolved by the {\it Fermi}-LAT {\it and} the putative population of MSPs that would explain the GCE. 
This is interesting to check as, if it were not possible to simultaneously fit both populations of MSPs with the same luminosity function as claimed by ref.~\cite{HooperMohlabeng2016}, then this would have implications for the bulge MSP population posited to explain the GCE. 
%
%
To presage our main result, we have found that the GCE provides only a  weak constraint on the luminosity function and that parameters can be found where both sets of data can be fit by the same luminosity function.
%
%
We modeled cases where the GCE is assumed to have a spherical geometry and where it is correlated with the stellar X-bulge. 
In Section \ref{methods}  our Bayesian Markov Chain Monte Carlo (MCMC) based approach to the problem is described. 
Our results are given in Section \ref{results}. 
Comparison to results from previous studies are discussed in Section \ref{discussion} and our conclusions are given in Section \ref{conclusions}. 
Some more technical aspects of our study are available in the Appendix.

\section{Methods}
\label{methods}

We used $71$ MSPs found within the Fermi-LAT third source catalog (3FGL) \cite{3FGL}. In the 3FGL, MSPs are not distinguished from pulsars and so the Galactic disk MSPs were identified by searching the Australia Telescope National Facility (ATNF) pulsar catalog \cite{ATNF:catalog} for 3FGL pulsars with periods of less than $10$ milliseconds and which were not associated with globular clusters. A small number of additional MSPs were identified through the use of an online list and found in the 3FGL catalog\footnote{https://confluence.slac.stanford.edu/display/GLAMCOG/Public+List+of+LAT-Detected+Gamma-Ray+Pulsars}. The list of the MSPs that we used is given in table \ref{tab:msp_data}.

\subsection{Modeling a Population of Millisecond Pulsars} \label{ssec:modelling_MSPs}

To simulate the GCE and a population of observed MSPs to compare with data, the underlying population of MSPs must be modeled. This requires spatial and luminosity models and, for the MSPs around the GC that may be responsible for the excess, a distribution of spectra.

As in ref.~\cite{HooperMohlabeng2016}, we used
a lognormal  luminosity function.
The lognormal distribution is one in which the logarithm of luminosity is normally distributed, its probability density function is:
\begin{equation}
p(L) = \frac{1}{\sigma_L L \sqrt{2 \pi}} \exp(\frac{-(\ln(L) - \ln(L_{\rm med}))^2}{2 \sigma_L^2})
\label{eq:luminosityfunction}
\end{equation}
\noindent where $L$ is luminosity, $L_{\rm med}$ is the median luminosity, and $\ln(L_{\rm med})$ and $\sigma_L$ are the mean and standard deviation of the normal distribution in $\ln(L)$. 

The spatial distribution of MSPs was divided into two components. One of these components models a population of MSPs scattered throughout the Milky Way disk according to the following density distribution \cite{Faucher10}:
\begin{equation}
\label{eq:rho_disk}
\rho_{\rm disk} (r_{\rm cyl}, z, N_{\rm disk}) = \frac{N_{\rm disk}}{4 \pi \sigma_r^2 z_0} \exp(-r_{\rm cyl}^2/2\sigma_r^2) \exp(-\abs{z}/z_0)
\end{equation}
\noindent where $r_{\rm cyl}$ is the distance from the GC projected onto the Galactic plane, $z$ is the distance perpendicular to the
Galactic plane, $N_{\rm disk}$ is the total number of MSPs in the entire Galactic disk, and $\sigma_r$ and $z_0$ are scale parameters. We take the distance from us to the GC to be 8.25kpc.

 The second component of the spatial distribution models the bulge population of MSPs potentially contributing to the GCE. In our region of interest, this density distribution has been found to be fitted by a spherically symmetric power law profile  \citep{Goodenough2009gk,hooperlinden2011,AbazajianKaplinghat2012,GordonMacias2013,MaciasGordon2014,Abazajian2014,Daylan:2014,CaloreCholisWeniger2015,Ajello2016,Dmitri2017}. In this article we use the parameterization
\begin{equation}
\label{eq:rho_bulge}
\rho_{\rm bulge} (r, N_{\rm bulge}) = \frac{3 N_{\rm bulge}}{20 \pi r_{\rm bulge}^{0.6}} r^{-2.4}, \quad 0 \le r < r_{\rm bulge}
\end{equation}
\noindent 
where 
$N_{\rm bulge}$ parameter is the total number of bulge MSPs in the Galaxy, $r$ is the distance from the GC, and $r_{\rm bulge}$ is the maximum radial extent of the bulge. The power law index is only known to about 20\% accuracy, but  our results are insensitive to this variation. We used a bulge radius of $r_{\rm bulge} = 3.1$ kpc to be consistent with ref.~\citep{HooperMohlabeng2016}. This is also the value determined from the COBE-DIRBE NIR maps  \cite{Mezger1996}.
\begin{figure} [t]
\centering
\includegraphics[width=1.0\linewidth]{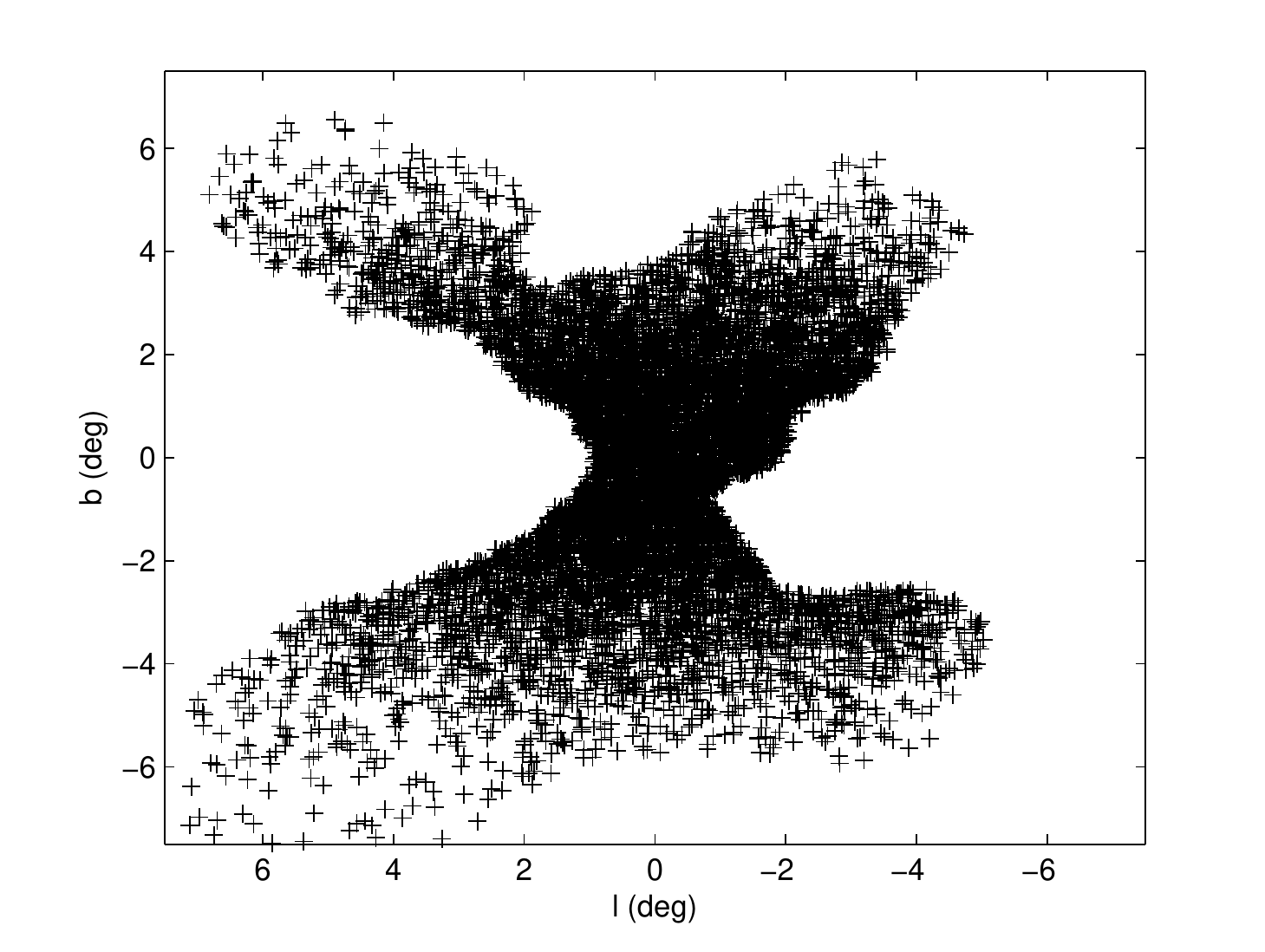}
\caption[X-shaped bulge spatial distribution]{Simulated spatial distribution of MSPs in X-shaped bulge.
}
\label{fig:Simulated_X_Bulge_Model}
\end{figure}

In addition to the spherically symmetric bulge model, an alternative model in which the bulge is X-shaped \citep{Macias2016} was also investigated. In this case, the spatial distribution is the product of the density profiles along the principle axes as shown in figure~$7$ of ref.~\cite{Bland-Hawthorn2016}, but includes only points randomly generated inside the projected X-shaped structure in Galactic coordinates. This X-shaped bulge distribution is shown in figure~\ref{fig:Simulated_X_Bulge_Model}.

Once the positions and luminosities of a population of MSPs (in both the disk and bulge) were simulated, an observed population could be found by applying a detection threshold based on the flux of each MSP, where the relationship between flux $F$, luminosity $L$ and distance $d$ is:

\begin{equation}
\label{eq:flux_luminosity}
F = \frac{L}{4 \pi d^2}
\end{equation}
\noindent A resolved MSP is one for which $F \ge F_{\rm th}$, where $F_{\rm th}$ is the threshold flux. The second Fermi-LAT catalog of gamma-ray pulsars \citep{FermiPulsarsCatalog} included an attempt to find the detection threshold as a function of $l$ and $b$ by adding simulated point sources at different positions in the sky and determining what the minimum flux needed was before they were detected. As the sensitivity to point sources depends strongly on the gamma-ray background, the detection threshold is highest near the Galactic plane.
As in ref.~\cite{HooperMohlabeng2016}, our modeled threshold did not solely depend on location in Galactic coordinates. Instead, $F_{\rm th}$ was  drawn from a lognormal distribution for each simulated MSP:
\begin{equation}
\label{eq:flux_threshold_pdf}
p(F_{\rm th}) = \frac{1}{\sigma_{\rm th} F_{\rm th} \sqrt{2 \pi}} \exp\left[\frac{-(\ln(F_{\rm th}) - (\mu_{\rm th}(l, b) + K_{\rm th}))^2}{2 \sigma_{\rm th}^2}\right]
\end{equation}

\noindent where $K_{\rm th}$ and $\sigma_{\rm th}$ are parameters, and $\mu_{\rm th}(l, b)$ is the natural logarithm of the threshold flux at $l$ and $b$ according to the pulsar catalog. A map of $2\exp(\mu_{\rm th}(l, b))$ is given in figure~16 of ref.~\cite{FermiPulsarsCatalog}.

The authors of ref.~\cite{FermiPulsarsCatalog} point out that these reported detection thresholds are likely to be underestimates; $K_{\rm th}$ is included as a parameter to account for this. The purpose of drawing $F_{\rm th}$ from the lognormal distribution is to approximate the variation that may occur due to uncertainty in the estimated threshold, or characteristics specific to individual pulsars, such as their spectra or light curves. The consequence of this is that the detection probability of a pulsar increases with flux, and does so particularly rapidly around the median of the threshold distribution, $\mu_{\rm th}(l, b) + K_{\rm th}$.

To simulate the GCE, a distribution of spectra must be modeled for the bulge MSPs. In the 3FGL catalog sources were fitted with three different spectral shapes. One of these is a power law:

\begin{equation}
\label{eq:power_law_dN_dE}
\frac{\dd{N}}{\dd{E}} \propto \left(\frac{E}{E_0}\right)^{-\Gamma}
\end{equation}

\noindent The second is an exponentially cutoff power law:

\begin{equation}
\label{eq:exp_cutoff_power_law_dN_dE}
\frac{\dd{N}}{\dd{E}} \propto \left(\frac{E}{E_0}\right)^{-\Gamma} \exp(-\frac{E}{E_{\rm cut}})
\end{equation}

\noindent Finally, a log parabolic spectrum was fitted for a small number of the observed MSPs:
\begin{equation}
\label{eq:log_parabolic_dN_dE}
\frac{\dd{N}}{\dd{E}} \propto \left(\frac{E}{E_0}\right)^{-\omega - \beta \ln(E/E_0)}
\end{equation}
 Each simulated bulge MSP was assigned the spectral shape and best fit parameters of a random resolved MSP. The proportionality constant was then found by requiring the energy integral over the spectrum (from $0.1$ to $100$ GeV) be equal to the flux of the simulated MSP:
\begin{equation}
\label{eq:E_dN_dE_integral}
F = \int_{0.1 \text{ GeV}}^{100 \text{ GeV}} E \frac{dN}{dE} dE
\end{equation}
 The simulated gamma-ray excess was then the sum of the spectra of all the MSPs in the relevant region of interest. This region is the $7^\circ \times 7^\circ$ box around the Galactic Center in the case of the spherically symmetric bulge, and for the X-shaped bulge it is the entire simulated bulge (which is inside a $15^\circ \times 15^\circ$ region).

\subsection{Markov Chain Monte Carlo} \label{ssec:MCMC}

In Subsection \ref{ssec:modelling_MSPs} a model was described which can produce a simulated population of MSPs, decide which are resolved, and simulate a gamma-ray excess based on the bulge MSP population. To find the parameters which may best reproduce the data, it is necessary to have a method for randomly sampling an arbitrary and potentially complex probability distribution of any number of dimensions.
The adaptive Metropolis MCMC algorithm described in ref.~\cite{Haario01} was used for this.
We used an unbinned Poisson distribution for the likelihood of the resolved MSPs \citep{cash1979}:
\begin{equation}
\label{eq:likelihood_unbinned}
\mathcal{L_{\rm res}} \propto \exp(-\lambda_{\rm res}) \prod_{i = 1}^{N} \rho(l_i,b_i, d_i, F_i)
\end{equation}
\noindent where $N=71$ is the number of resolved MSPs,  $\lambda_{\rm res}$ is the expected number of resolved MSPs, $\rho(l_i,b_i, d_i, F_i)$ is the modeled density of resolved MSPs at Galactic coordinates $l_i,b_i$, distance $d_i$, and flux $F_i$.

We used the following parameters $(\theta$):
\begin{enumerate}
\item The total expected number of observed MSPs $\lambda_{\rm res}$ from eq.~(\ref{eq:likelihood_unbinned}).
\item The natural log of the ratio of total number of disk to bulge MSPs $\ln(r_{\rm d/b})=\ln(N_{\rm disk}/N_{\rm bulge})$ from eqs.~(\ref{eq:rho_bulge}) and (\ref{eq:rho_disk}).
\item Luminosity function parameters $\log_{10}(L_{\rm med})$ and $\sigma_L$ for the lognormal distribution given by eq.~(\ref{eq:luminosityfunction}).
\item The flux threshold distribution parameters $K_{\rm th}$ and $\sigma_{\rm th}$ from eq.~(\ref{eq:flux_threshold_pdf}).
\item The spatial model parameters $\sigma_r$ and $z_0$.
\item  The distance parameter for each observed MSP, $d_i$.
\item The parallax distance measurement probability parameter $\alpha$.
\item The flux of each observed MSP, $F_i$.
\item The flux threshold at each observed MSP, $F_{{\rm th},i}$.
\end{enumerate}
Note that $\lambda_{\rm res}$ appears as a parameter (rather than being fixed to the observed number of MSPs) because the observed number is essentially drawn from a Poisson distribution with an unknown expected value.  The parameters $\lambda_{\rm res}$ and $\sigma_{\rm th}$ were required to be greater than $0$ and $\sigma_L$ was restricted to values above $0.5$. A lower limit of $0.8$ for $\sigma_L$ is justified in ref.~\cite{HooperMohlabeng2016} by considering the luminosity distribution of those MSPs with parallax distance measurements. Here, those measurements were included as priors on the distance parameters corresponding to those MSPs. The upper limit for $\sigma_r$ was $10.0$, this was chosen as it is expected to be approximately $5$ kpc \citep{Story:2007xy,Faucher10,Gregoire:2013yta,Lorimer2013,HooperMohlabeng2016}.
Prior boundaries were also used for other parameters, these boundaries were located in places where 
the likelihood 
was very low (i.e.\ the proposed set of parameters would be extremely unlikely to be accepted).
Table~\ref{tab:PriorBoundaries} lists prior boundaries placed on each of the parameters.

\begin{table} [!htb]
\centering
\begin{tabular}{|c|c|c|}
\hline
Parameter & Lower Boundary & Upper Boundary \\ \hline \hline
$\lambda_{\rm res}$ & $0.0$ & - \\ \hline
$\ln(r_{\rm d/b})$ & - & - \\ \hline
$K_{\rm th}$ & $0.0$ & $3.9$ \\ \hline
$\sigma_{\rm th}$ & $0.0$ & $2.3$ \\ \hline
$\sigma_r/{\rm kpc}$ & $0.0$ & $10.0$ \\ \hline
$z_0/{\rm kpc}$ & $0.0$ & $2.0$ \\ \hline
$d_i/{\rm kpc}$ & $0.0$ & - \\ \hline
$\log_{10}(L_{\rm med}/({\rm erg \cdot s}^{-1}))$ & $29.0$ & $38.0$ \\ \hline
$\sigma_L$ & $0.5$ & $3.5$ \\ \hline
$\alpha/{\rm kpc}$ & $0.01$ & $10.0$ \\ \hline 
$F_{{\rm th},i}/({\rm erg\cdot cm^{-2}\cdot s^{-1}})$&0&$F_i$\\ \hline
$F_i/({\rm erg\cdot cm^{-2}\cdot s^{-1}})$
& $\exp(\ln(L_{\rm med})-6\sigma_L)/ (4\pi d_i^2)$ & $\exp(\ln(L_{\rm med})+6\sigma_L)/ (4\pi d_i^2)$\\ \hline
\end{tabular}

\caption[Model parameter boundaries]{Prior boundaries used for each parameter in the MCMC simulation for a lognormal  luminosity function.}
\label{tab:PriorBoundaries}
\end{table}

We do not use the dispersion estimates of the MSPs distances as they may have a high systematic error \citep{FermiPulsarsCatalog,HooperMohlabeng2016}.

The expected excess that would be produced by the bulge MSPs can be found by multiplying $N_{\rm bulge}$ by the expected contribution of a single MSP. For the X-shaped bulge GCE data, some of the higher energy bins had large relative errors, so they were combined. The large size of the resulting bin meant that it was necessary to take $\left(\frac{\dd{N}}{\dd{E}}\right)_{{\rm sim,}i}$ to be the mean across the bin, $N/(E_{{\rm max},i} - E_{{\rm min},i})$, where $E_{{\rm min},i}$ and $E_{{\rm max},i}$ are the edges of the bin. The highest energy bin we used in the X-bulge case was
($28$~GeV~$\le E\le 	158$~GeV).
It  corresponded to an expected value of around 43 counts which was the smallest value for both the spherical and X-bulge spectrum. So as the counts in each bin are reasonably high, the GCE likelihood will be well approximated \cite{Berry-Esseen} by a Gaussian:
\begin{equation}
\label{eq:likelihood_GCE}
\mathcal{L}_{\rm GCE} \propto \prod_{i = 1}^{N} \exp(-\left(\left(\frac{\dd{N}}{\dd{E}}\right)_{{\rm sim,}i} - \left(\frac{\dd{N}}{\dd{E}}\right)_{{\rm data,}i}\right)^2 /(2 \sigma_{{\rm data,}i}^2))
\end{equation}
\noindent where $\left(\frac{\dd{N}}{\dd{E}}\right)_{{\rm sim,}i}$ and $\left(\frac{\dd{N}}{\dd{E}}\right)_{{\rm data,}i}$ are respectively the simulated and observed gamma-ray excess with uncertainty $\sigma_{{\rm data,}i}^2$ at $E_i$. This is calculated using the GCE spectra
from refs.~\cite{MaciasGordon2014} and \cite{Macias2016}.

%
%
%

The second component of the likelihood $\mathcal{L}_{\rm res}$ is  given by eq.~(\ref{eq:likelihood_unbinned}) 
with 
\begin{equation}
\label{eq:density_res_F_i}
\rho(l_i,b_i, d_i, F_i)\propto (\rho_{\rm disk}+\rho_{\rm bulge}) p(L_i) 
d_i^4
\end{equation}
where $p(L_i)$ is the luminosity function given by eq.~(\ref{eq:luminosityfunction}), 
 the factor $d_i^4$ is from a product of a factor of $d_i^2$  from the Jacobian for the change of variables $F_i$ to $L_i$ and another factor of $d_i^2$ from the Jacobian for the change of variables
from Cartesian to Galactic coordinates as $\rho_{\rm disk}$ and $\rho_{\rm bulge}$ are densities in Cartesian coordinates.
Also, $\rho_{\rm disk}$ and $\rho_{\rm bulge}$ are determined by eqs.~(\ref{eq:rho_disk}) and (\ref{eq:rho_bulge}) respectively. For our prior distribution we used:
\begin{equation}
\label{eq:prior}
p(\theta) \propto \prod_{i = 1}^{N}
p(F_{{\rm th},i})
p(F_i \vert F_{{\rm data,}i}, \sigma_{{\rm data,}i}) \prod_{i\in {\rm parallax}} p(d_i)
\end{equation}
where  $p(F_{{\rm th},i}) $ is given by eq.~(\ref{eq:flux_luminosity}), $p(F_i \vert F_{{\rm data,}i} ,\sigma_{{\rm data,}i})$ is a normal distribution with the observed MSP's flux and uncertainty (obtained from 3FGL) as the mean and standard deviation respectively
and $p(d_i)$ is constructed for the subset of MSPs which had parallax measurements using the  best fit value and errors.

As parallax measurements are more likely to be available for nearer MSPs, a third component of the likelihood was used. This $\mathcal{L}_{\rm parallax}$ is the product of modeled probabilities of observed MSPs having or not having a parallax distance measurement:
\begin{equation}
\mathcal{L}_{\rm parallax} = \prod_{i \in {\rm parallax}} \exp(-d_i / \alpha) \prod_{i \not\in {\rm parallax}} (1 - \exp(-d_i / \alpha))
\end{equation}
The posterior distribution was obtained by combining the likelihood and prior distributions:
\begin{equation}
p(\theta\vert {\rm data})\propto \mathcal{L}_{\rm res} \times \mathcal{L}_{\rm GCE} \times \mathcal{L}_{\rm parallax} \times p(\theta)
\end{equation}
We marginalized over the $F_{{\rm th},i}$ and the $F_i$ variables using numerical integration. 
 The other variables were sampled using the MCMC.
Twelve Markov chains were constructed of five million iterations each for the spherical and X-shaped models.

Although, we have a large number of parameters that we are fitting, this is ameliorated as the flux $(F_i)$ and distance parameters $(d_i)$ are all linked by the geometric model (eqs.~(\ref{eq:rho_disk}) and (\ref{eq:rho_bulge}) or the X-bulge version) and the luminosity function given by eq.~(\ref{eq:luminosityfunction}). Also, the threshold parameters $(F_{{\rm th},i})$ are linked by the threshold prior giving by eq.~(\ref{eq:flux_threshold_pdf}). This regularizes the problem in a similar way to ridge regression \cite{Gelman13}.

We tested our model fits using posterior predicted p-values (see Chapter 6 of ref.~\cite{Gelman13}). For every 1250th step in the Markov chains we produced a set of simulated resolved MSPs and GCE data points. The resolved MSPs were binned in longitude, latitude, flux and distance. In order to determine if a point was well fit we evaluated the posterior predicted p-value as the fraction of simulated points of greater magnitude than the data point. The fit was considered to be acceptable if the posterior predicted p-value was between 0.025 and 0.975.

To further test our overall model fits, we randomly pick 200 parameter sets from the Markov chains, and for each of these we simulate 500 sets of resolved MSP and GCE data. We bin the resolved MSPs in $l$, $b$ and $\log(F)$, we take the mean number in each bin to be the expectation value and use Poisson distributions to calculate the likelihood for the observed data as well as each set of simulated data:
\begin{equation}
\mathcal{L} = \prod_{i = 1}^{n} \frac{\lambda_i^{N_i} \exp(-\lambda_i)}{N_i!}
\end{equation}
where $n$ is the total number of bins, $N_i$ is the number of resolved MSPs in bin $i$ with expectation value $\lambda_i$. We used 5 bins along each of $l$ and $b$, and 7 along $\log(F)$ for a total of 175 bins. We can then calculate a p-value using the fraction of likelihoods for simulated data sets lower than that of the observed data. Similarly, we can also find p-values for the GCE using eq.~(\ref{eq:likelihood_GCE}) where $\left(\frac{\dd{N}}{\dd{E}}\right)_{{\rm sim,}i}$ is the mean of the simulated data. The simulated data in each bin is also shifted by a random sample from a normal distribution with a standard deviation equal to the error on the observed data. The resulting distributions of p-values indicate the extent to which the distribution of model parameters in the Markov chains fits the data.





\section{Results}
\label{results}

For both the spherical and X-bulge models, corner plots are presented \cite{corner} showing the results of the MCMC simulations. These figures show histograms of the two parameters associated with the number of MSPs ($\lambda_{\rm res}$ and $\ln(r_{\rm d/b})$), the luminosity function parameters, the flux threshold distribution parameters, the spatial parameters and the parallax model parameter. In addition to those histograms, these figures display the distributions for each pair of model parameters along with $68\%$ and $95\%$ contours. The distance parameters ($d_i$) are not shown as these were considered nuisance parameters. 

To check the fit quality,
for each chain, the parameters of a few thousand evenly spaced points were used to generate a simulated set of resolved MSPs. This simulated data was binned in $l$, $b$, $\log_{10}(F)$ and $\log_{10}(d)$ and the means and standard deviations of each bin are compared to the observed data in a set of figures for each model. The distribution of the simulated GCE is also plotted along with the measured data. A pair of plots are also shown which display the distribution of the number of simulated resolved MSPs both inside and outside the projected bulge region compared with the observations. The projected bulge region is all Galactic coordinates where the probability of a bulge MSP being modeled is non-zero.

We also estimated how many bulge MSPs are expected to be resolved by Fermi-LAT and future experiments 
with double and four times the sensitivity of the current Fermi-LAT data.
We did this by evaluating, for every step in the Markov chains, the expected number of observed bulge MSPs along with the expected number for the cases where the flux thresholds were divided by factors of two and four. Using the series of expected values for each of the three cases, Poisson distributions were randomly sampled giving a discrete distribution of values $N$, the number of resolved bulge MSPs. From these three distributions, the overall probability for getting any $N$ can be estimated. Histograms of these probability distributions are shown for each model along with a table showing the mean and the probability of $N>0$.

\FloatBarrier
\subsection{
Spherically Symmetric Bulge} \label{ssec:ResultsLognormalSpherical}

\begin{figure} [!htb]
\centering
\includegraphics[width=1.0\linewidth]{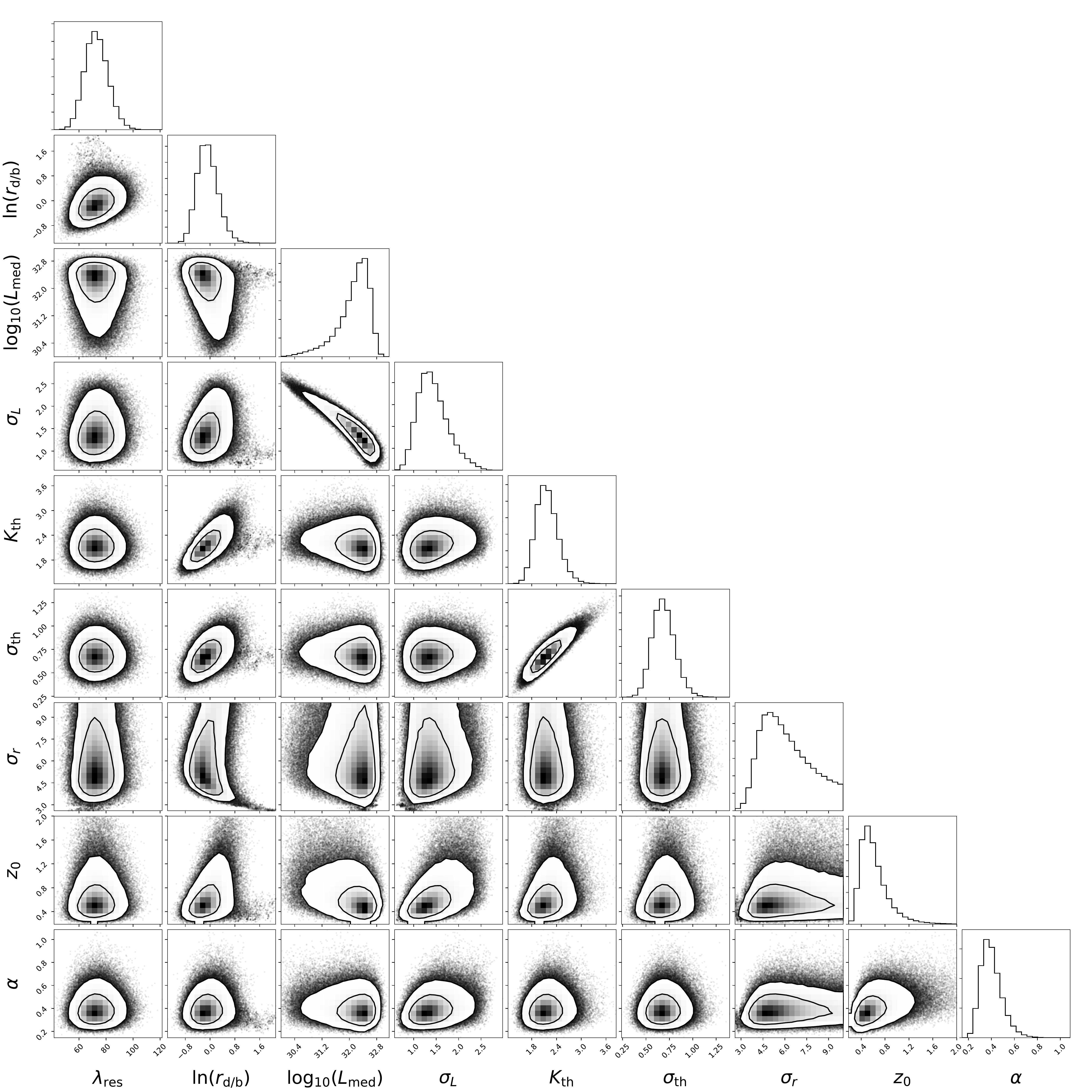}
\caption[Results for lognormal luminosity distribution and spherical bulge model]{Distribution of points in Markov chains for the lognormal luminosity distribution and spherical bulge model. Contours contain 68\% and 95\% of points. The units of $L_{\rm med}$ and $\alpha$ are $({\rm erg \cdot s}^{-1})$ and kpc respectively.}
\label{fig:CornerLognormalSpherical}
\end{figure}

In this section the results for the 
 spherically symmetric bulge model 
are presented. 
In figure~\ref{fig:CornerLognormalSpherical}, the distribution of points in the set of Markov chains produced for this model is shown. Correlations can be seen between $\log_{10}(L_{\rm med})$ and $\sigma_L$ as well as between the three parameters $\ln(r_{\rm d/b})$, $K_{\rm th}$ and $\sigma_{\rm th}$. Table \ref{tab:ResultsLognormalSpherical} presents the mean and error for each model parameter.

\begin{table} [!htb]
\centering
\begin{tabular}{|c|c|c|}
\hline
Parameter & Mean & Error \\ \hline \hline
$\lambda_{\rm res}$ & $73$ & $9$ \\ \hline
$\ln(r_{\rm d/b})$ & $-0.1$ & $0.3$ \\ \hline
$\log_{10}(L_{\rm med}/({\rm erg \cdot s}^{-1}))$ & $32.1$ & $0.5$ \\ \hline
$\sigma_L$ & $1.4$ & $0.3$ \\ \hline
$K_{\rm th}$ & $2.2$ & $0.3$ \\ \hline
$\sigma_{\rm th}$ & $0.7$ & $0.1$ \\ \hline
$\sigma_r/{\rm kpc}$ & $6$ & $2$ \\ \hline
$z_0/{\rm kpc}$ & $0.6$ & $0.2$ \\ \hline
$\alpha/{\rm kpc}$ & $0.43$ & $0.09$ \\ \hline
\end{tabular}

\caption[Mean values and 68\% confidence interval errors  for lognormal luminosity distribution and spherical bulge model parameters]{Mean values  and 68\% confidence interval errors for lognormal luminosity distribution and spherical bulge model parameters.}
\label{tab:ResultsLognormalSpherical}
\end{table}

\begin{figure} [!htb]
\centering
\subfigure{\centering\includegraphics[width=0.49\linewidth]{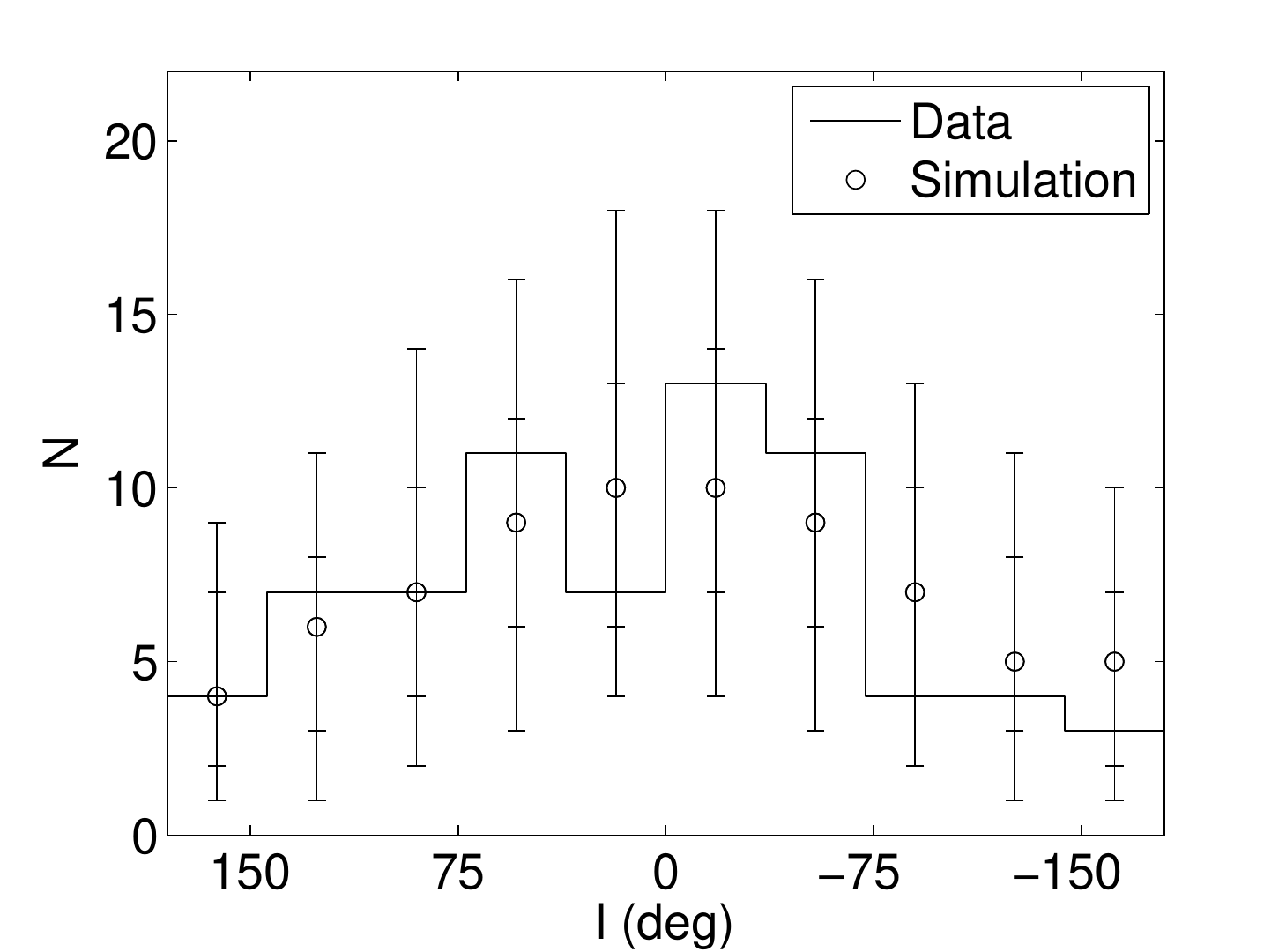}}
\subfigure{\centering\includegraphics[width=0.49\linewidth]{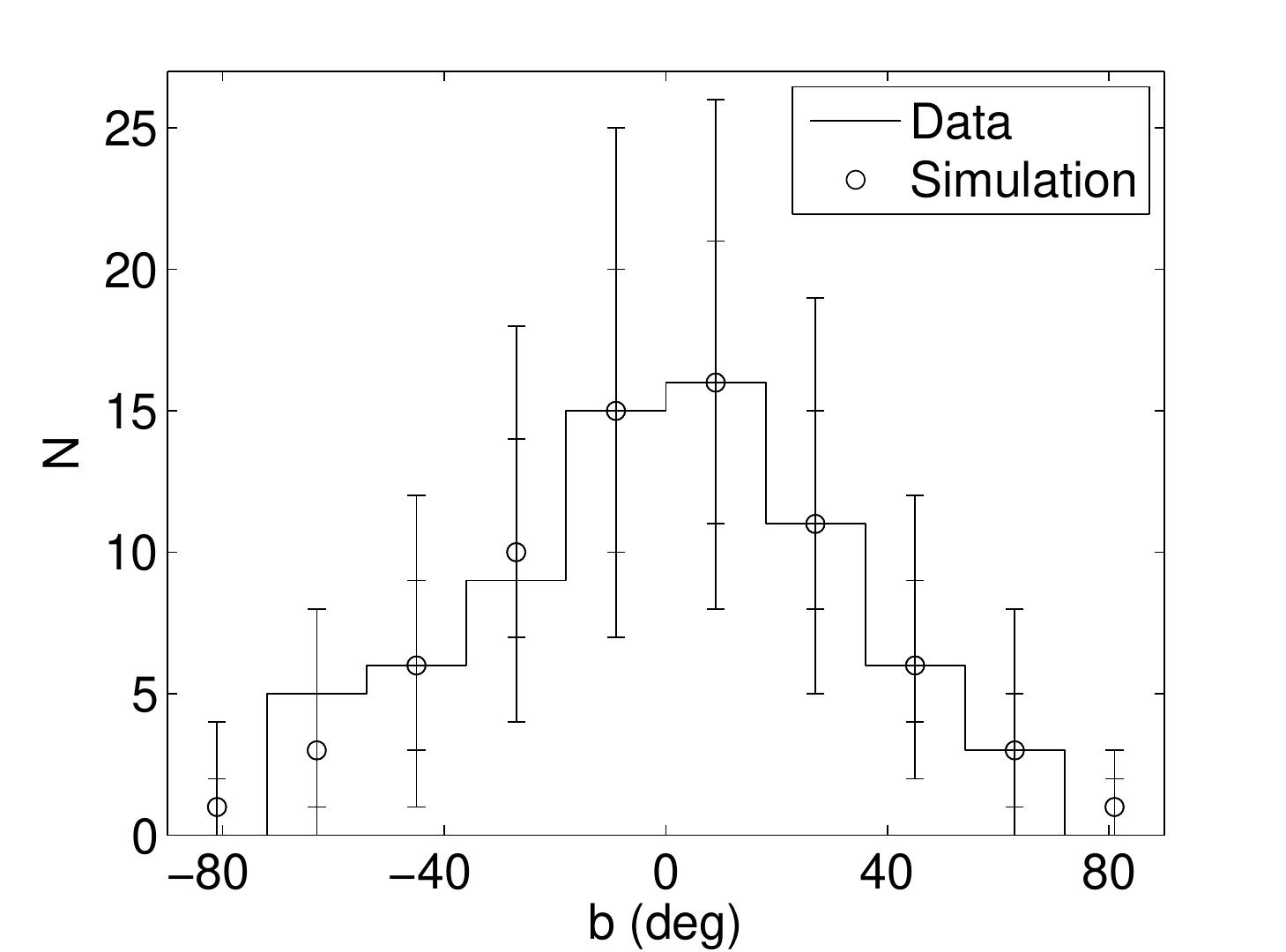}}
\subfigure{\centering\includegraphics[width=0.49\linewidth]{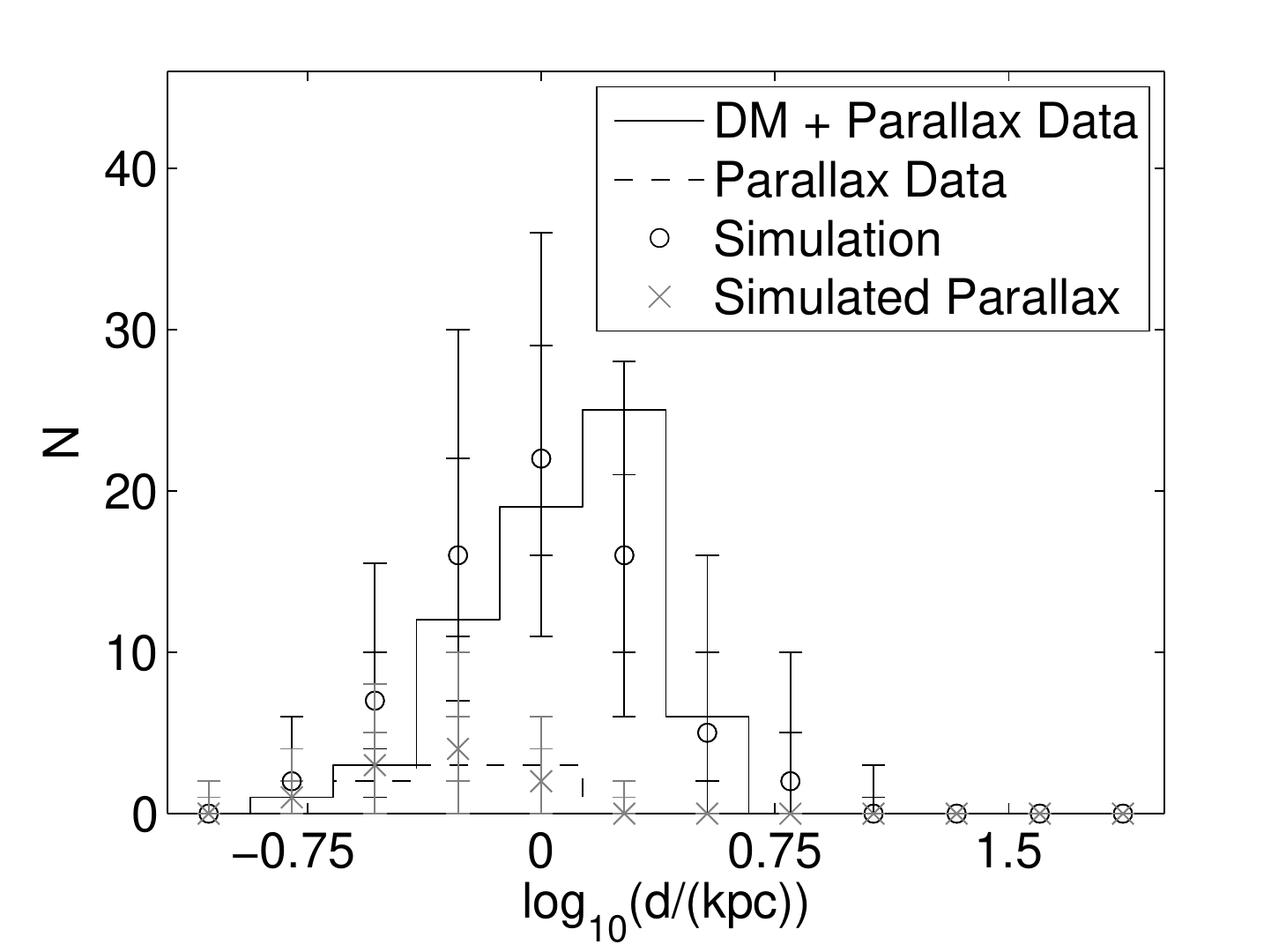}}
\subfigure{\centering\includegraphics[width=0.49\linewidth]{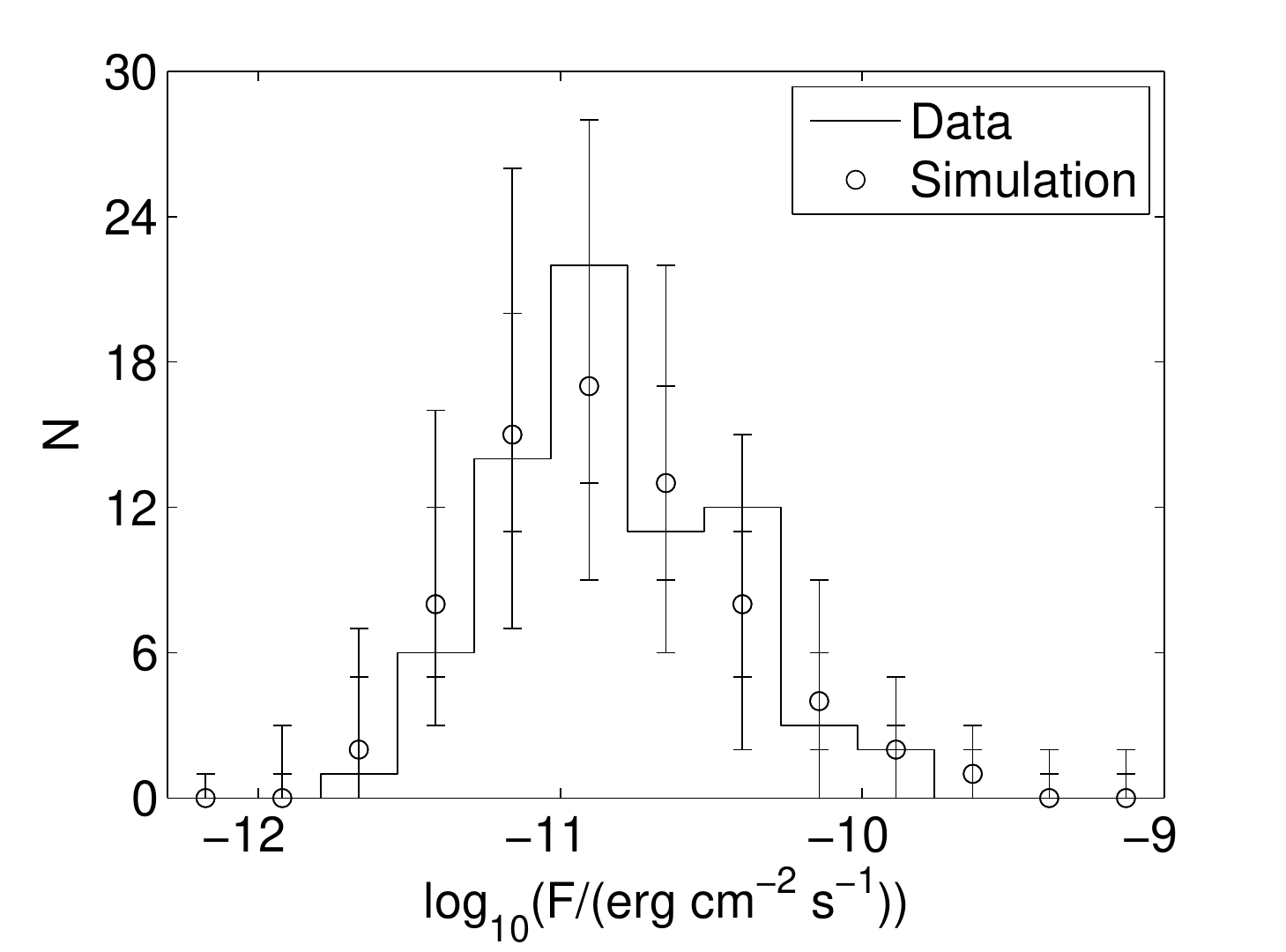}}
\subfigure{\centering\includegraphics[width=0.49\linewidth]{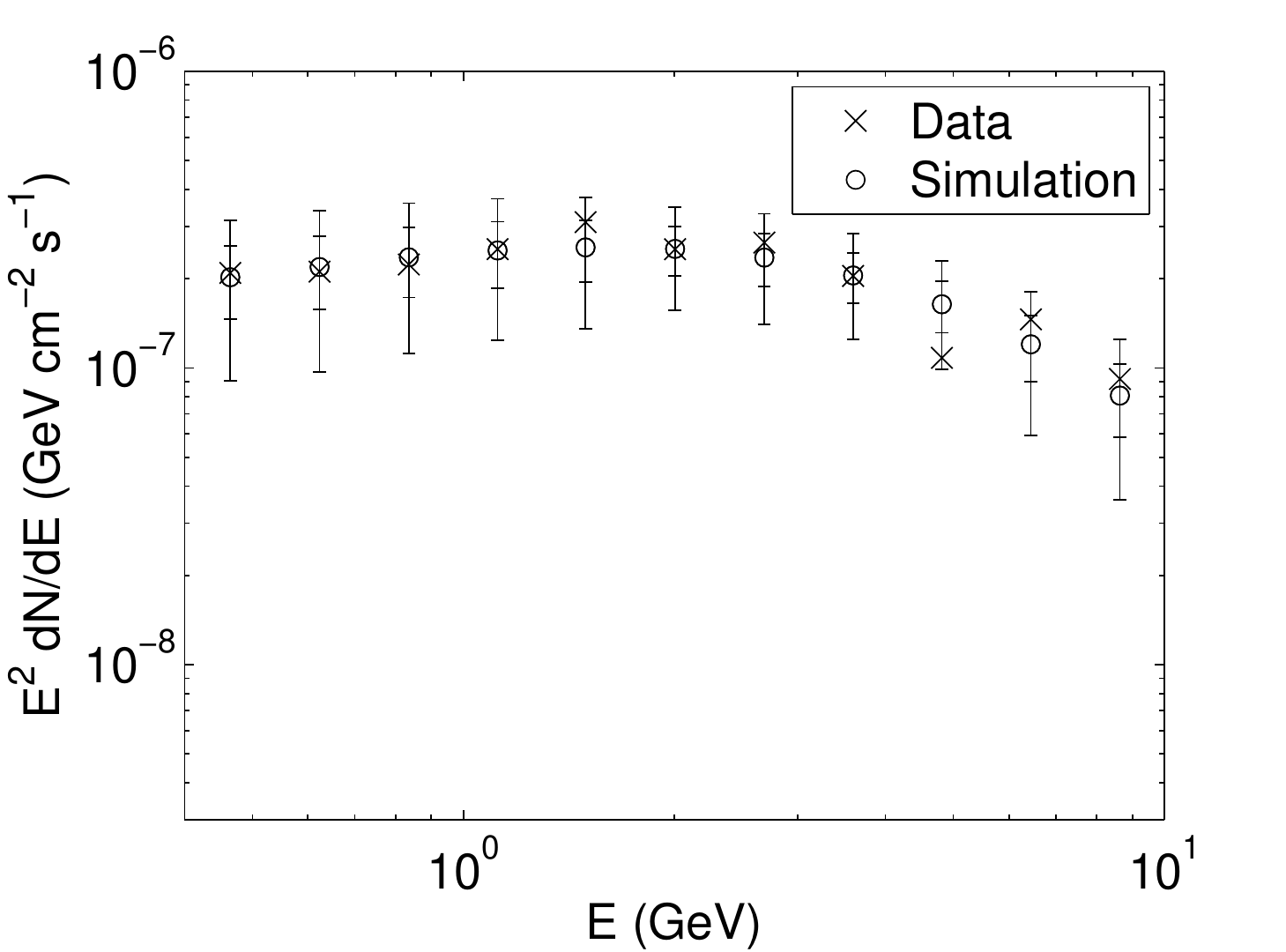}}
\caption[Simulated data for lognormal luminosity distribution and spherical bulge model and lognormal  luminosity function]{Simulated observed distribution of MSPs in longitude, latitude, distance and flux, and simulated GCE for the  spherical bulge model and lognormal  luminosity function. In the distance plot, DM means dispersion measure derived distances. The simulated points and error bars are the median, 68\% and 95\% intervals of the simulated populations in each bin. The error bars on the simulated GCE include the errors on the observed data. }
\label{fig:SimDataLognormalSpherical}
\end{figure}

\begin{figure} [!htb]
\centering
\subfigure{\centering\includegraphics[width=0.49\linewidth]{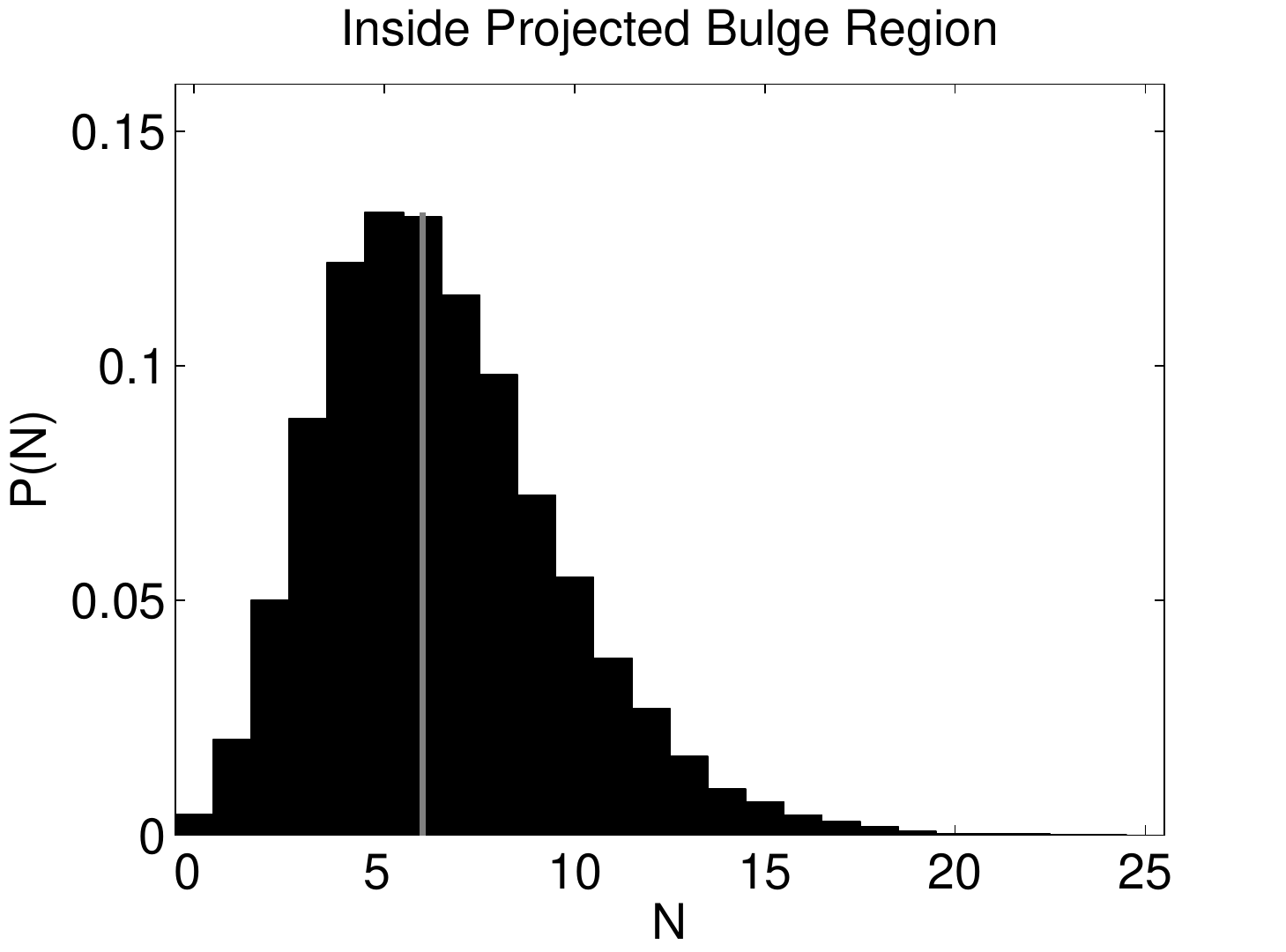}}
\subfigure{\centering\includegraphics[width=0.49\linewidth]{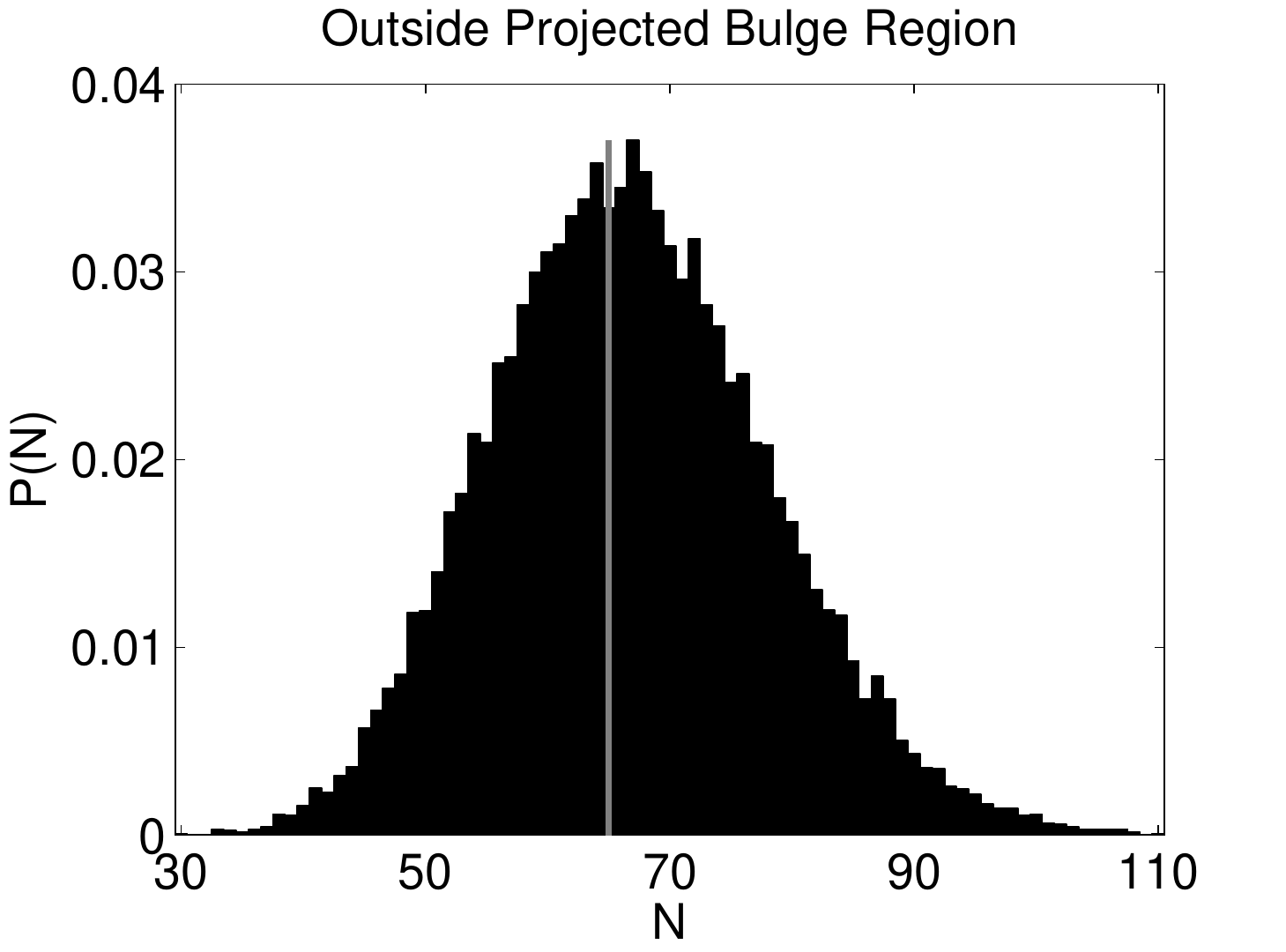}}
\caption[Probability of $N$ observed MSPs inside and outside the projected bulge using lognormal luminosity distribution and spherical bulge model]{The approximate probability distributions of observing $N$ MSPs inside and outside the projected bulge using the lognormal luminosity distribution and spherical bulge model. 
Here the projected bulge region is anywhere within 
21.3$^\circ$ of the GC.
The gray lines are the observed number for each case.}
\label{fig:NObsBulgeRegionLognormalSpherical}
\end{figure}

A set of figures show the results of using the sets of parameters in the Markov chains constructed using this model to simulate populations of MSPs. In figure~\ref{fig:SimDataLognormalSpherical}, the binned distributions of resolved MSPs in longitude, latitude, distance and flux are shown, as well as the simulated GCE produced by the bulge population. The distribution of the number of resolved MSPs inside and outside of the projected bulge region is seen in figure~\ref{fig:NObsBulgeRegionLognormalSpherical}.

\begin{figure} [!htb]
\centering
\includegraphics[width=0.9\linewidth]{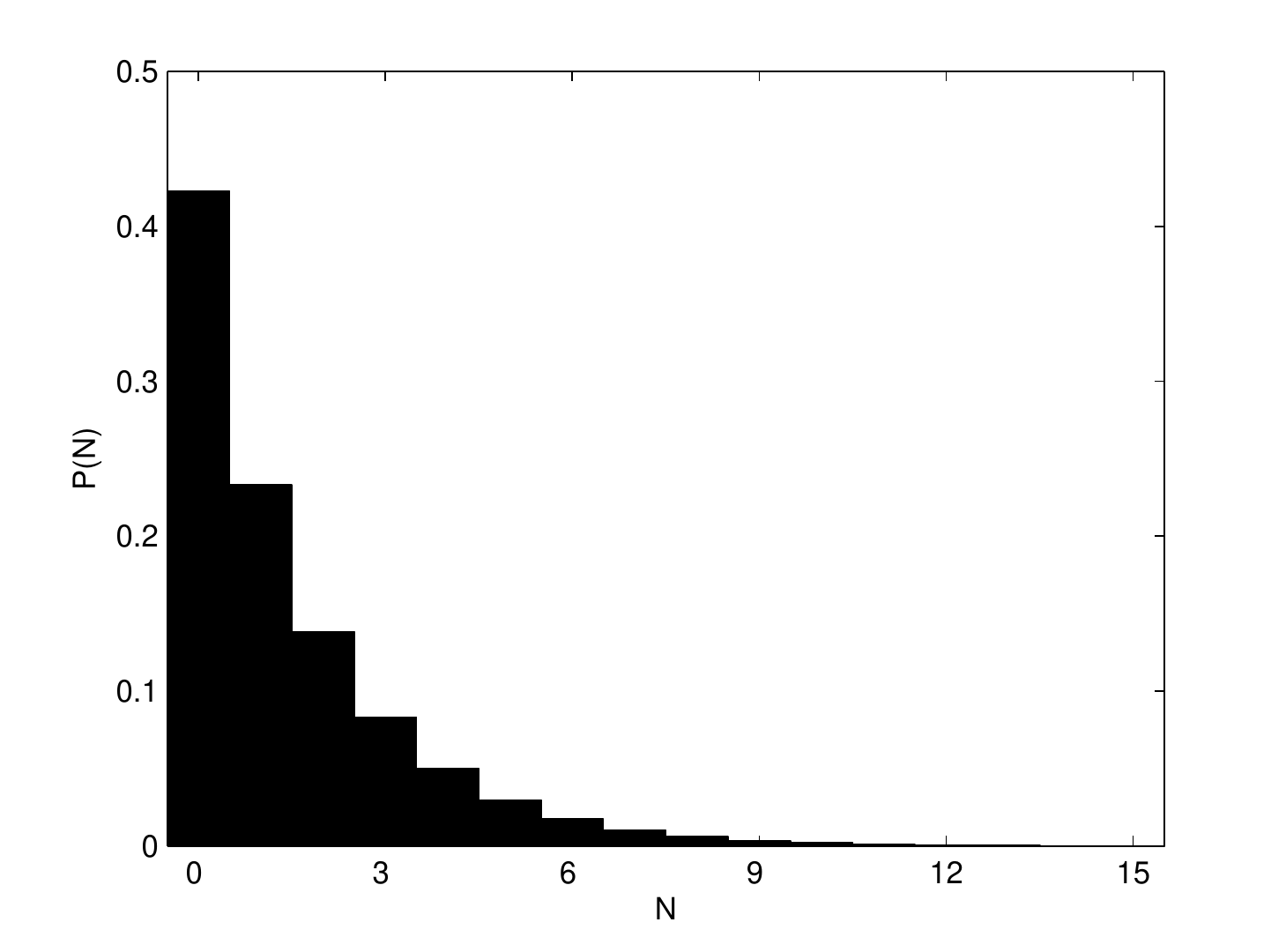}
\caption[Probability of $N$ observed bulge MSPs for lognormal luminosity distribution and spherical bulge model]{The probability distribution of observing $N$ MSPs from the bulge population based on the  spherical bulge model and lognormal  luminosity function.}
\label{fig:NObsLognormalSpherical}
\end{figure}

\begin{figure} [!htb]
\centering
\subfigure{\centering\includegraphics[width=0.49\linewidth]{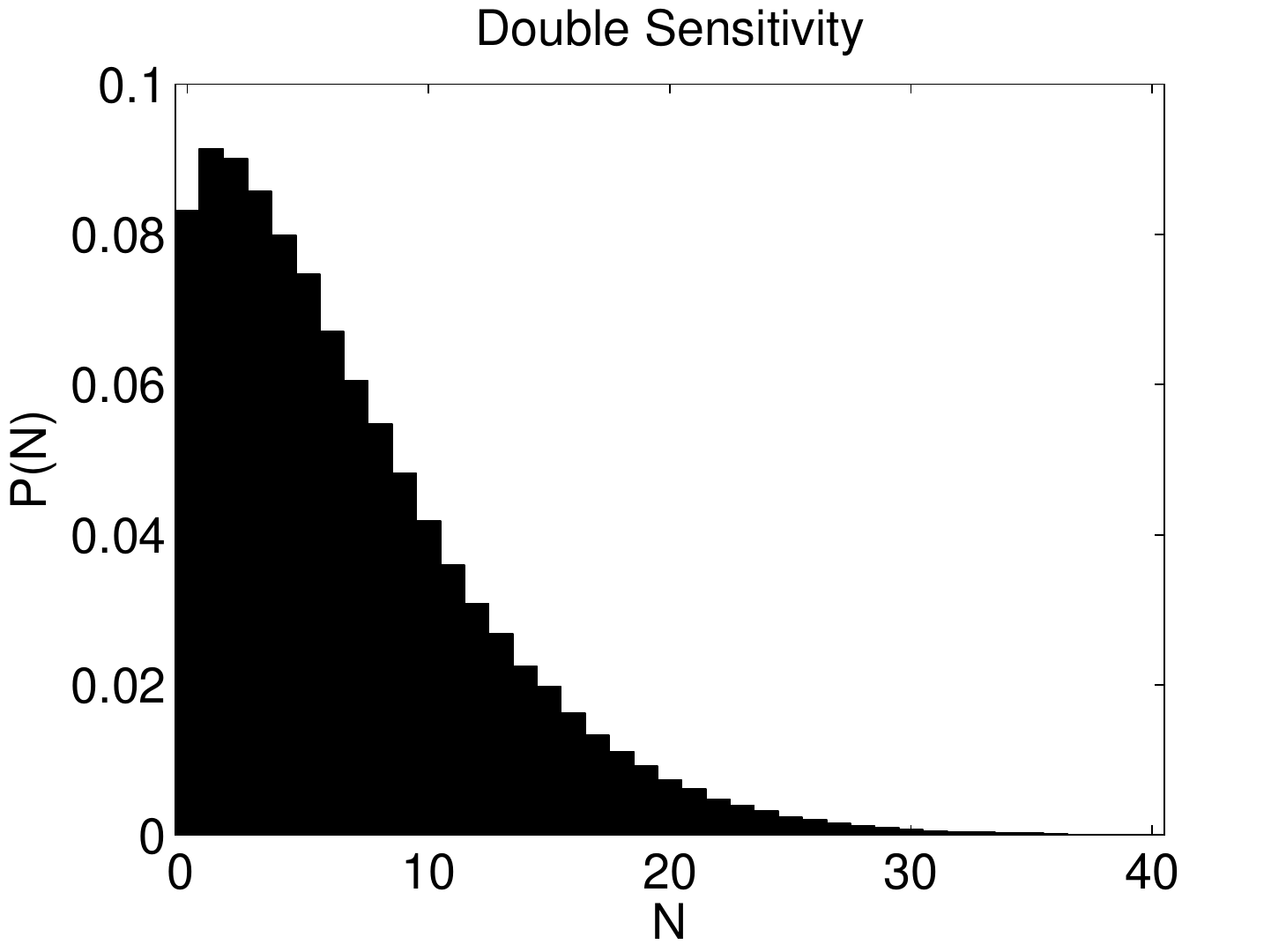}}
\subfigure{\centering\includegraphics[width=0.49\linewidth]{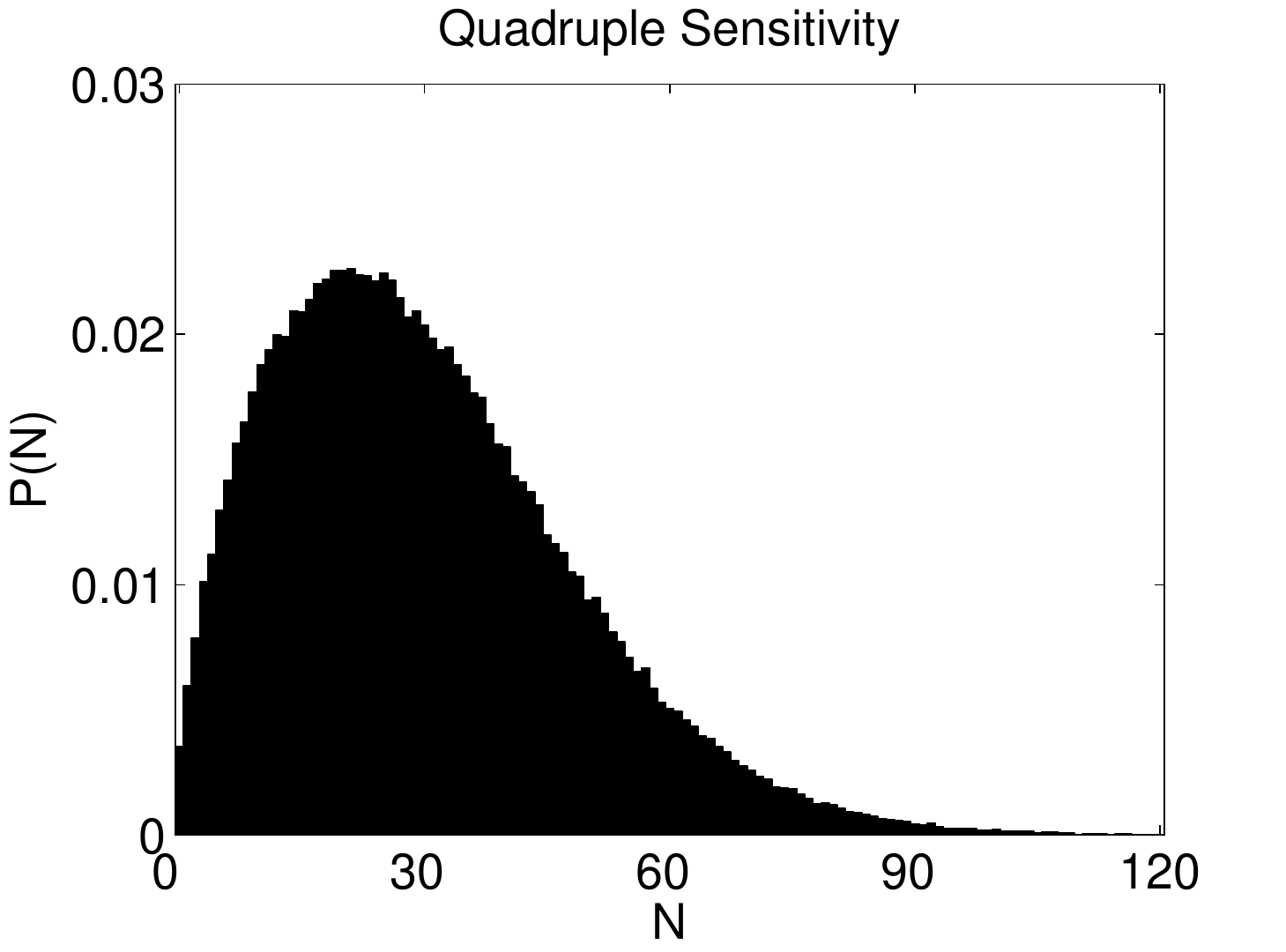}}
\caption[Probability of $N$ observed bulge MSPs for lognormal luminosity distribution and spherical bulge model with double or quadruple sensitivity]{The probability distribution of observing $N$ MSPs from the bulge population based on the lognormal luminosity distribution and spherical bulge model with double or quadruple sensitivity.}
\label{fig:NObsDoubleQuadrupleLognormalSpherical}
\end{figure}

\begin{figure} [!htb]
\centering
\subfigure{\centering\includegraphics[width=0.49\linewidth]{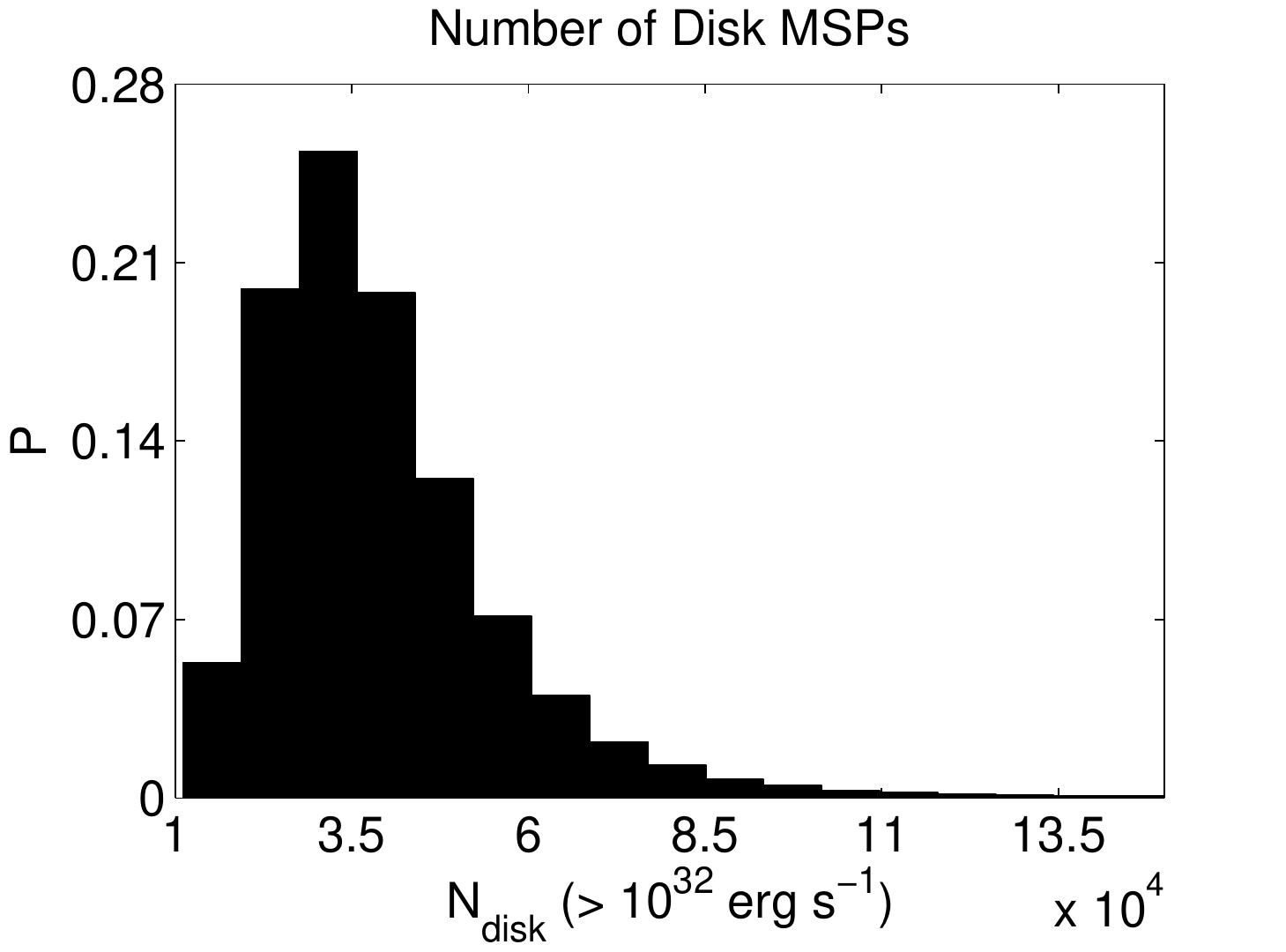}}
\subfigure{\centering\includegraphics[width=0.49\linewidth]{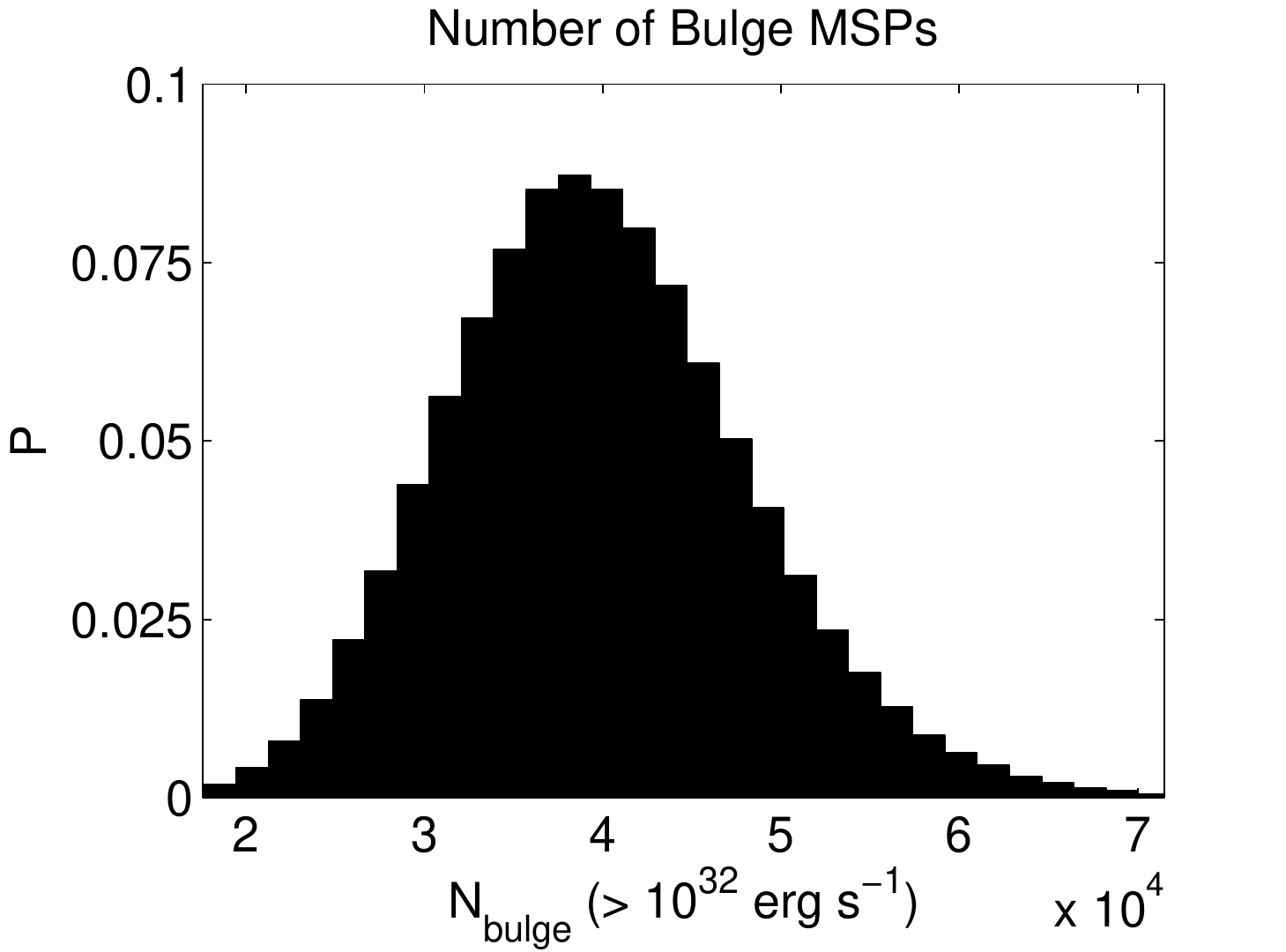}}
\caption[Number of MSPs for spherical bulge model and lognormal  luminosity function]{The distribution of the number of MSPs with luminosity greater than $10^{32}$ ${\rm erg \cdot s}^{-1}$ for the spherical bulge model and lognormal  luminosity function.}
\label{fig:NumberMSPsLognormalSpherical}
\end{figure}

Table \ref{tab:ObsBulgeMSPLognormalSpherical} shows the probability of observing any MSPs from the bulge population and the expected number based on the fitted flux threshold parameters as well as where the detection sensitivity has been doubled and quadrupled. These three probability distributions in the number of resolved bulge MSPs are shown in figures~\ref{fig:NObsLognormalSpherical} and \ref{fig:NObsDoubleQuadrupleLognormalSpherical}. The number of MSPs in the disk with luminosity greater than $10^{32}$ ${\rm erg \cdot s}^{-1}$ was $\left(4 \pm 2\right) \times 10^4$ and the number in the bulge was $\left(4.0 \pm 0.9\right) \times 10^4$. In figure~\ref{fig:NumberMSPsLognormalSpherical}, the distributions of the number of MSPs are shown.

\begin{table} [!htb]
\centering
\begin{tabular}{|c|c|c|}
\hline
Sensitivity Factor & Mean $N$ & P($N > 0$) \\ \hline \hline
1.0 & $1.42$ & $0.577$ \\ \hline
2.0 & $6.81$ & $0.917$ \\ \hline
4.0 & $30.1$ & $0.997$ \\ \hline
\end{tabular}

\caption[Probability of observing bulge MSPs and expected number for lognormal luminosity distribution and spherical bulge model]{Expected number of observed MSPs located in the bulge and probability of observing one or more for lognormal luminosity distribution and spherical bulge model. The sensitivity factor is the number of times more sensitive the prediction is than the current data.}
\label{tab:ObsBulgeMSPLognormalSpherical}
\end{table}

\FloatBarrier
\subsection{
X-shaped Bulge} \label{ssec:ResultsLognormalX}

\begin{figure} [!htb]
\centering
\includegraphics[width=1.0\linewidth]{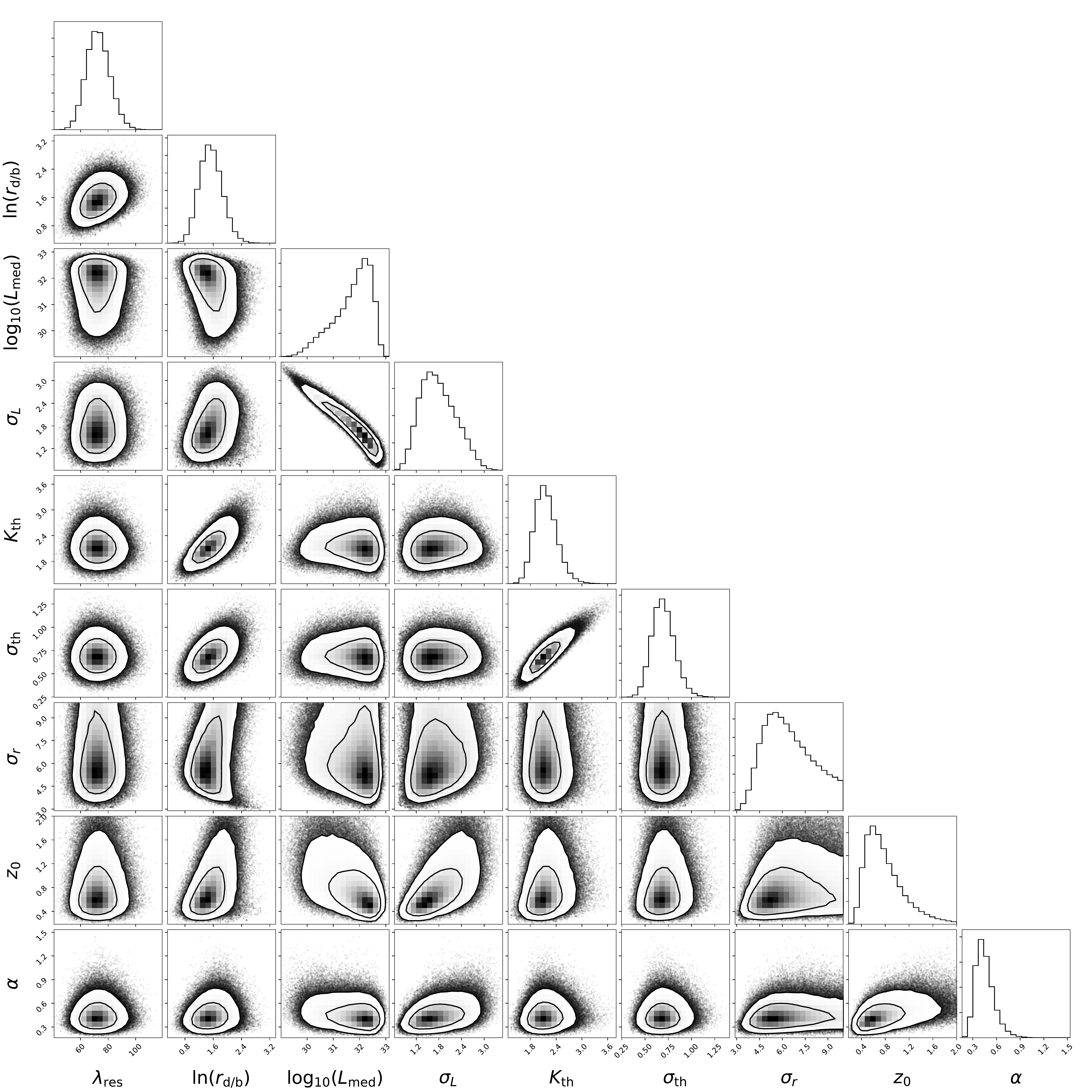}
\caption[Results for lognormal luminosity distribution and X-shaped bulge model]{Distribution of points in Markov chains for the lognormal luminosity distribution and X-shaped bulge model. Contours contain 68\% and 95\% of points. The units of $L_{\rm med}$ and $\alpha$ are $({\rm erg \cdot s}^{-1})$ and kpc respectively.}
\label{fig:CornerLognormalX}
\end{figure}

This section presents the results for the spatial model 
 which has an X-shaped bulge. The MCMC simulation results are shown in figure~\ref{fig:CornerLognormalX}. There are, as for the spherically symmetric bulge case, clear correlations between $\log_{10}(L_{\rm med})$ and $\sigma_L$ and between the three parameters $\ln(r_{\rm d/b})$, $K_{\rm th}$ and $\sigma_{\rm th}$. Means and errors for each of the model parameters are listed in table \ref{tab:ResultsLognormalX}.

\begin{table} [!htb]
\centering
\begin{tabular}{|c|c|c|}
\hline
Parameter & Mean & Error \\ \hline \hline
$\lambda_{\rm res}$ & $73$ & $9$ \\ \hline
$\ln(r_{\rm d/b})$ & $1.5$ & $0.3$ \\ \hline
$\log_{10}(L_{\rm med}/({\rm erg \cdot s}^{-1}))$ & $31.7$ & $0.7$ \\ \hline
$\sigma_L$ & $1.8$ & $0.5$ \\ \hline
$K_{\rm th}$ & $2.2$ & $0.3$ \\ \hline
$\sigma_{\rm th}$ & $0.7$ & $0.1$ \\ \hline
$\sigma_r/{\rm kpc}$ & $6$ & $2$ \\ \hline
$z_0/{\rm kpc}$ & $0.8$ & $0.3$ \\ \hline
$\alpha/{\rm kpc}$ & $0.5$ & $0.1$ \\ \hline
\end{tabular}

\caption[Mean values and  68\% confidence interval errors for lognormal luminosity distribution and X-shaped bulge model parameters]{Mean values and  68\% confidence interval errors for lognormal luminosity distribution and X-shaped bulge model parameters.}
\label{tab:ResultsLognormalX}
\end{table}

\begin{figure} [!htb]
\centering
\subfigure{\centering\includegraphics[width=0.49\linewidth]{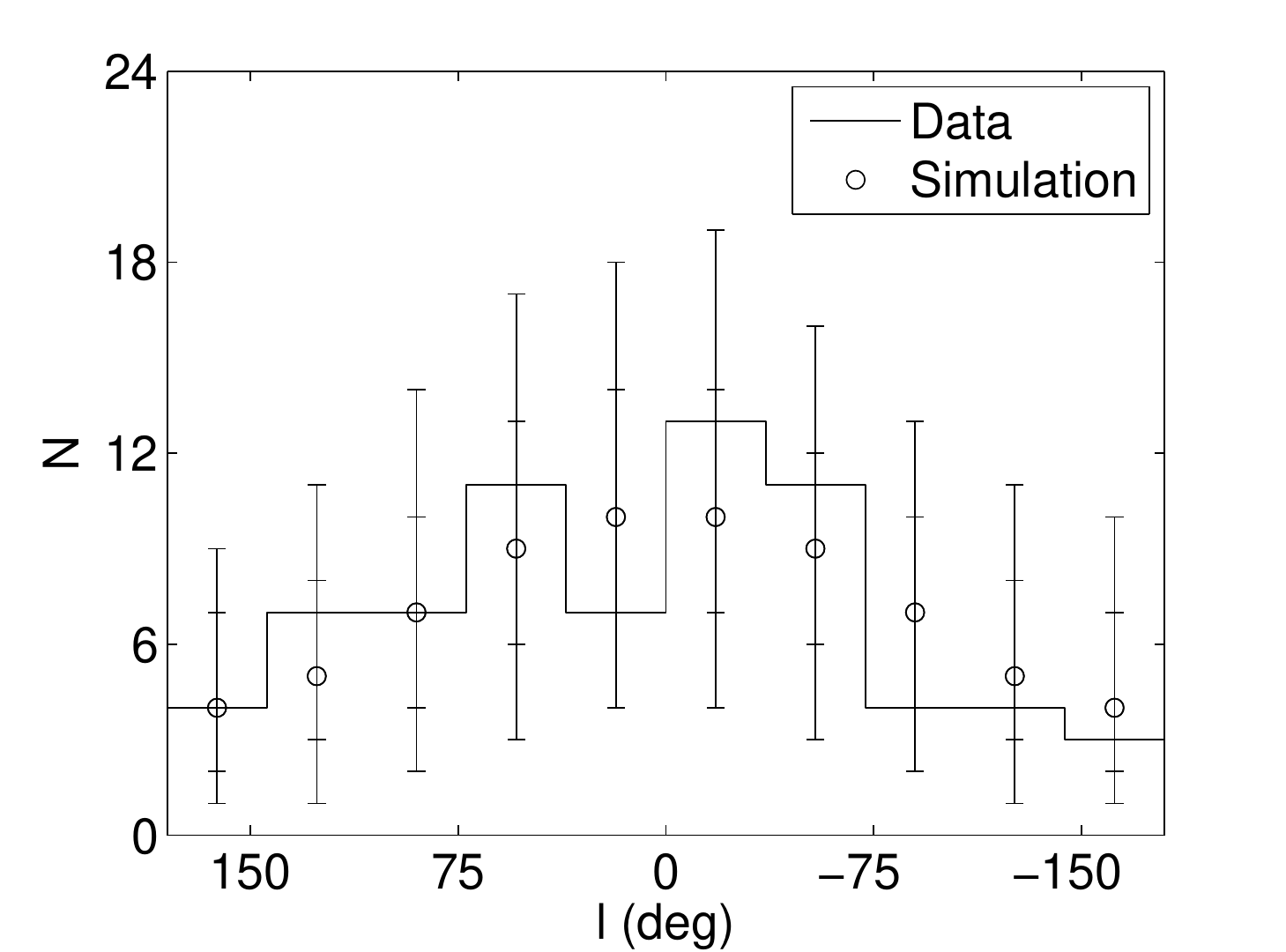}}
\subfigure{\centering\includegraphics[width=0.49\linewidth]{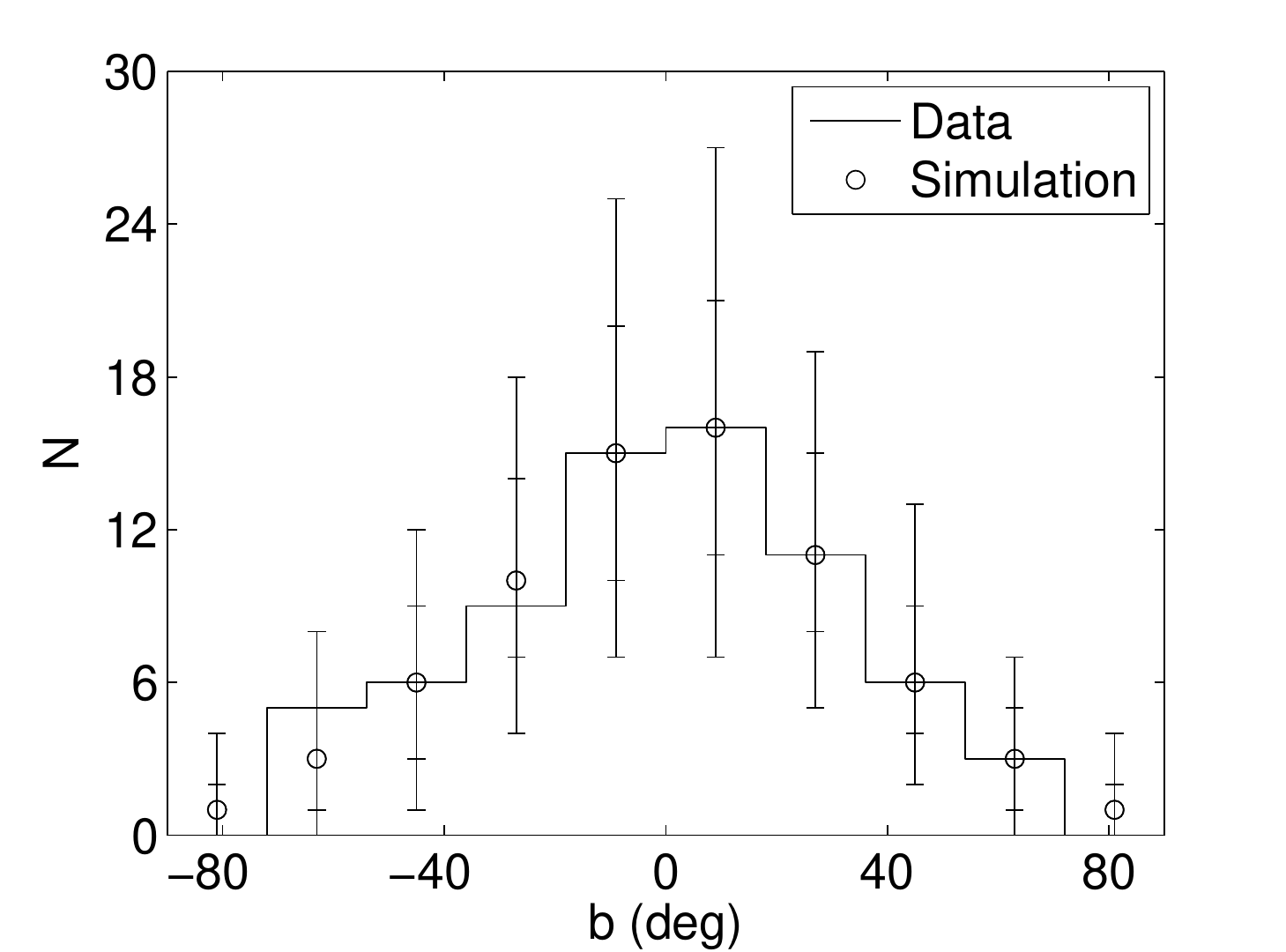}}
\subfigure{\centering\includegraphics[width=0.49\linewidth]{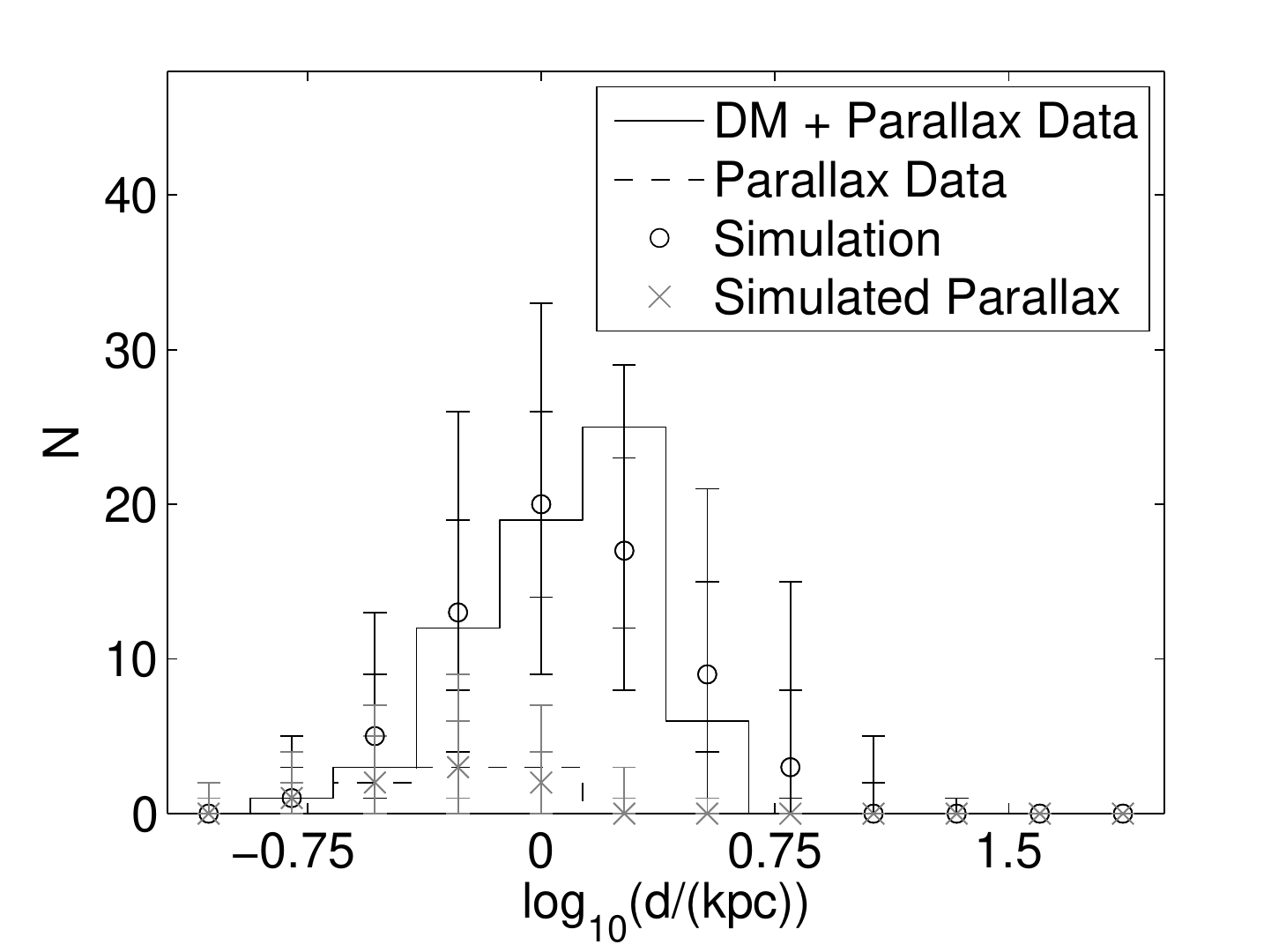}}
\subfigure{\centering\includegraphics[width=0.49\linewidth]{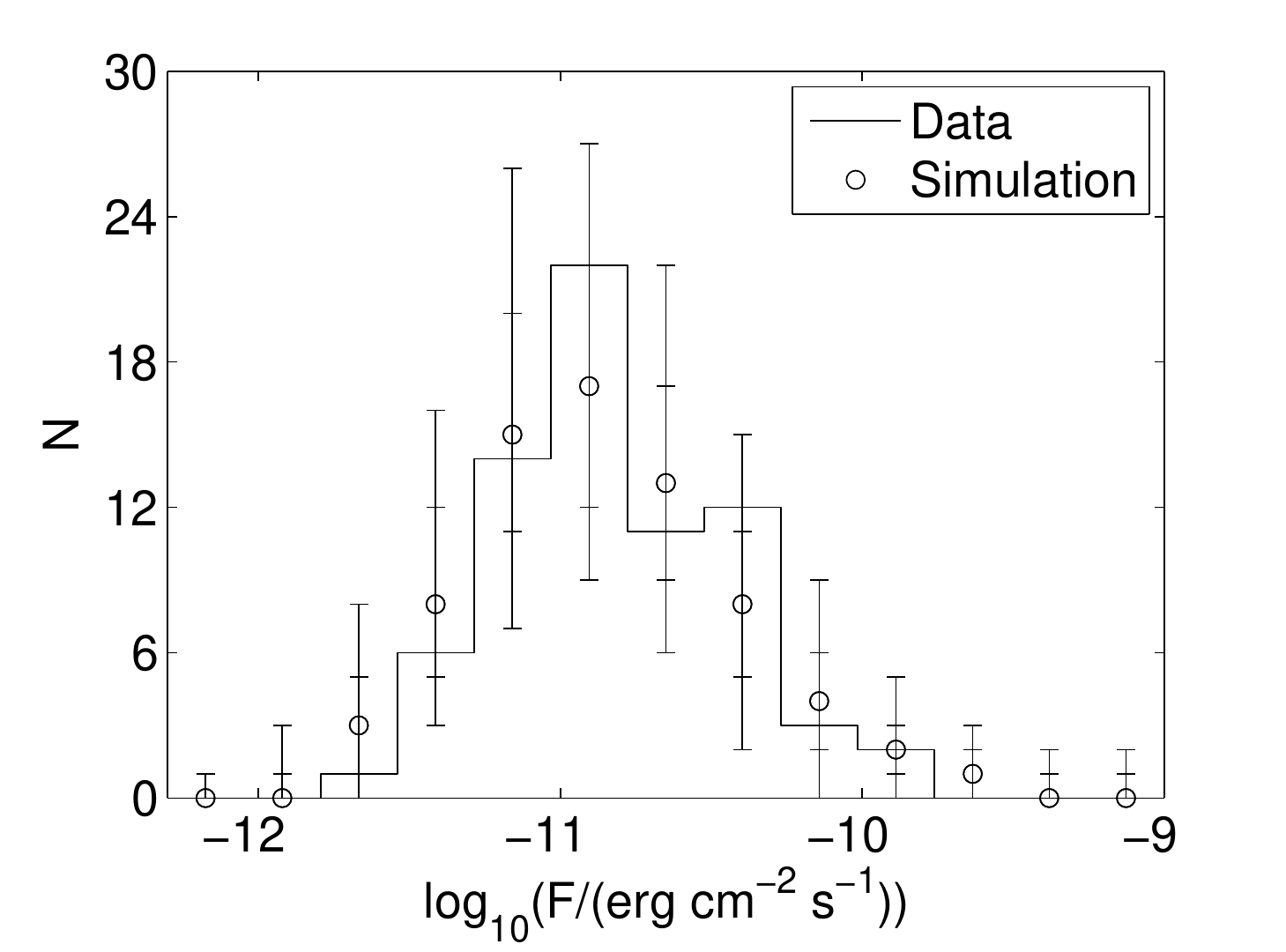}}
\subfigure{\centering\includegraphics[width=0.49\linewidth]{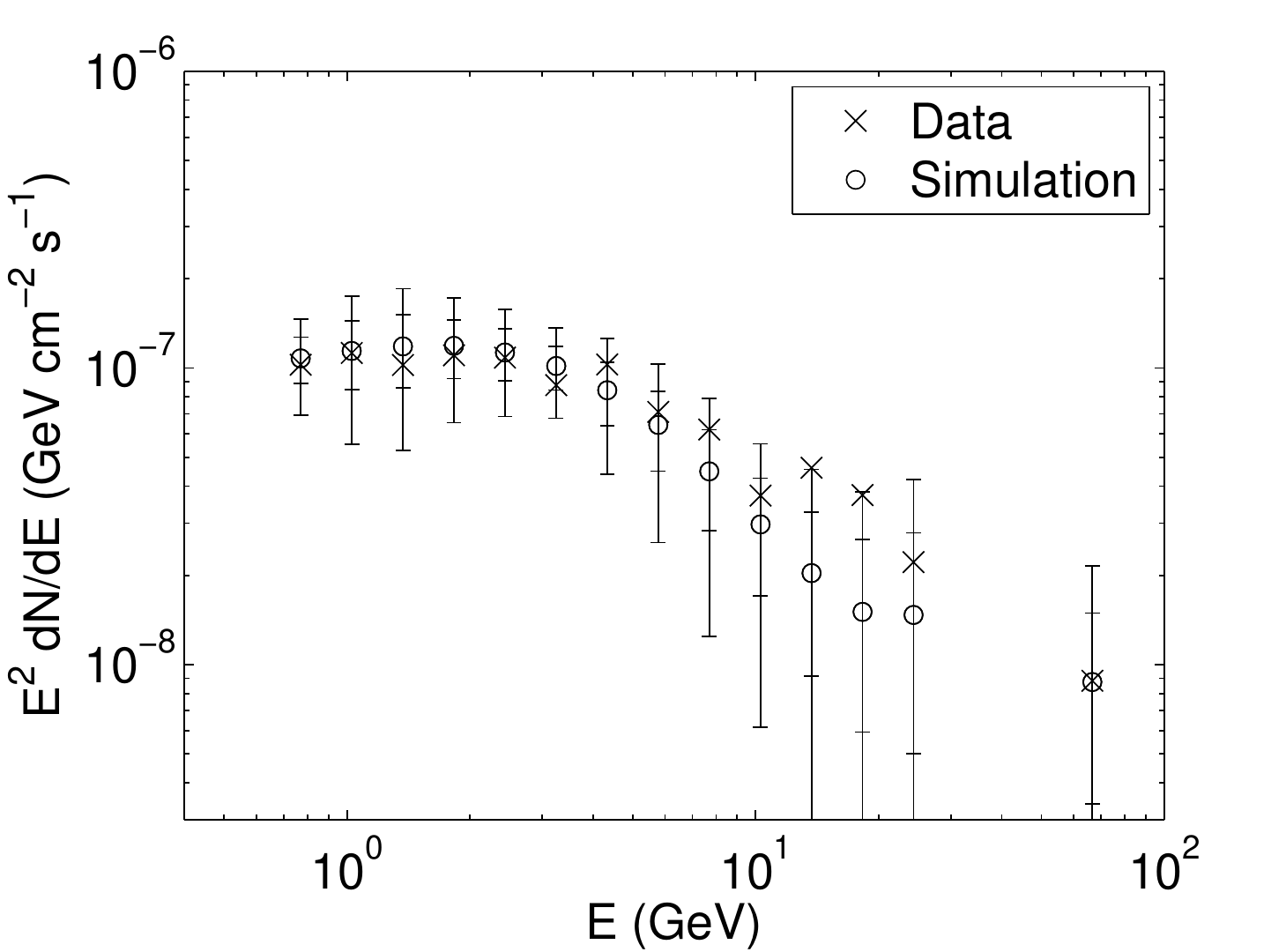}}
\caption[Simulated data for 
the X-shaped bulge model]{Simulated observed distribution of MSPs in longitude, latitude, distance and flux, and simulated Galactic Center excess for the lognormal luminosity distribution and X-shaped bulge model. In the distance plot, DM means dispersion measure derived distances. The simulated points and error bars are the median, 68\% and 95\% intervals of the simulated populations in each bin. The error bars on the simulated GCE include the errors on the observed data. Note that the last bin covers the $28$~GeV~$\le E\le 	158$~GeV range.
}
\label{fig:SimDataLognormalX}
\end{figure}

\begin{figure} [!htb]
\centering
\subfigure{\centering\includegraphics[width=0.49\linewidth]{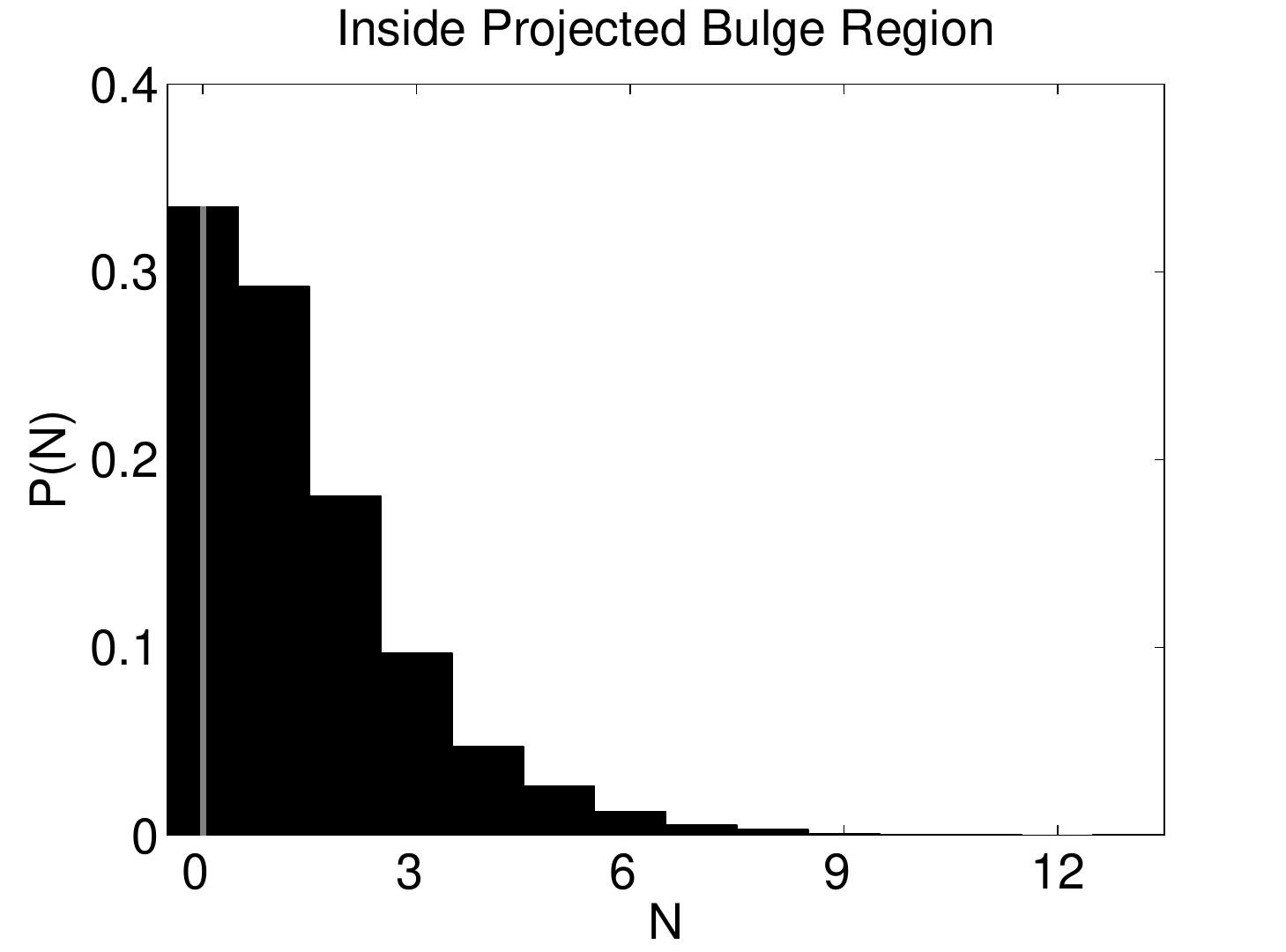}}
\subfigure{\centering\includegraphics[width=0.49\linewidth]{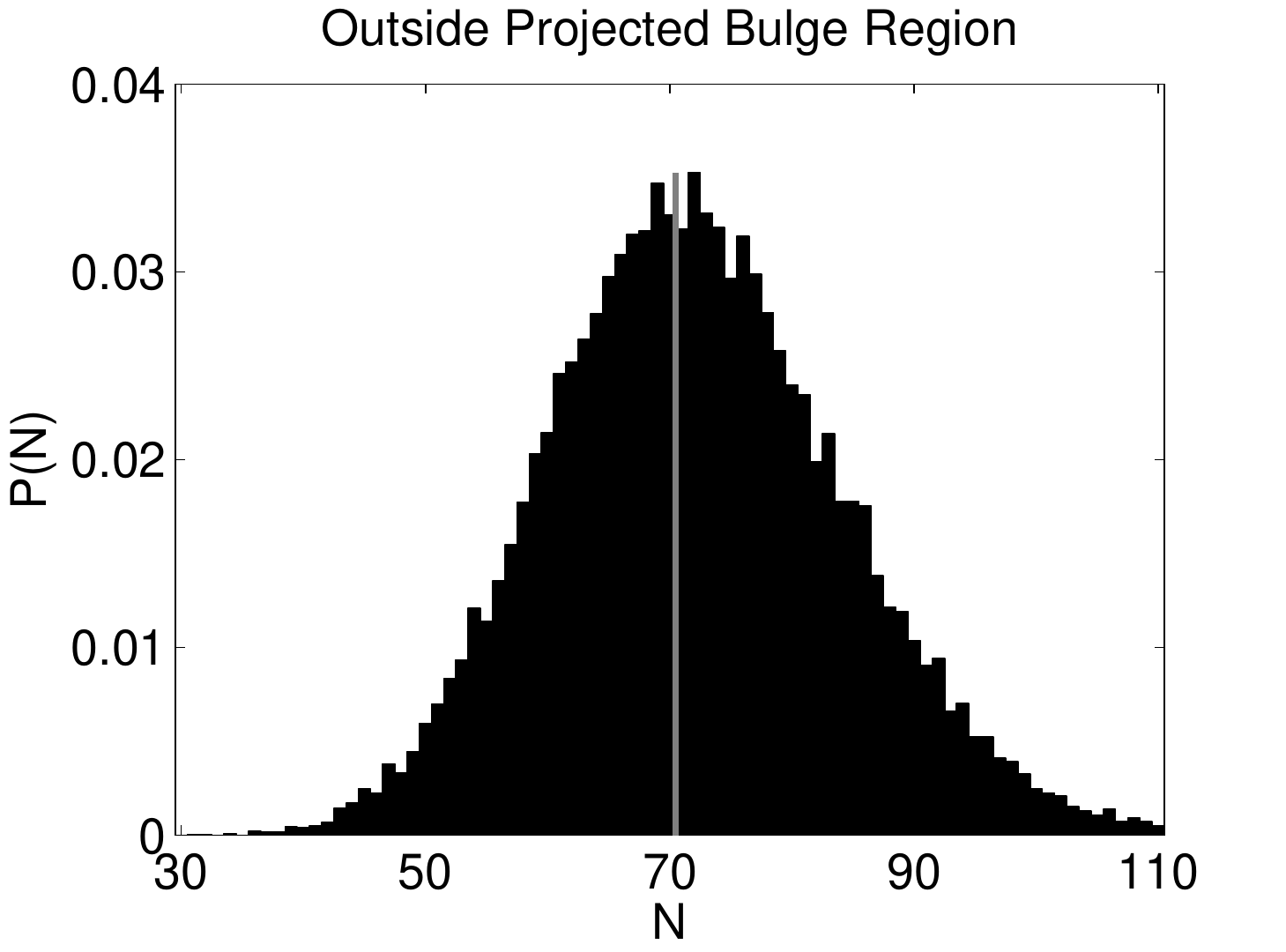}}
\caption[Probability of $N$ observed MSPs inside and outside the projected bulge using lognormal luminosity distribution and X-shaped bulge model]{The approximate probability distributions of observing $N$ MSPs inside and outside the projected bulge using the 
lognormal luminosity distribution and
X-shaped bulge model. 
Here the projected bulge region is the inner $15^\circ\times 15^\circ$ around the GC for which the X-bulge template of ref.~\cite{Macias16} is non-zero.
The gray lines are the observed number for each case.}
\label{fig:NObsBulgeRegionLognormalX}
\end{figure}

The simulated GCE and simulated distributions of resolved MSPs in longitude, latitude, distance and flux are shown in figure~\ref{fig:SimDataLognormalX}, with the distribution of the numbers located inside and outside the region of the projected bulge displayed in figure~\ref{fig:NObsBulgeRegionLognormalX}.

\begin{figure} [!htb]
\centering
\includegraphics[width=0.9\linewidth]{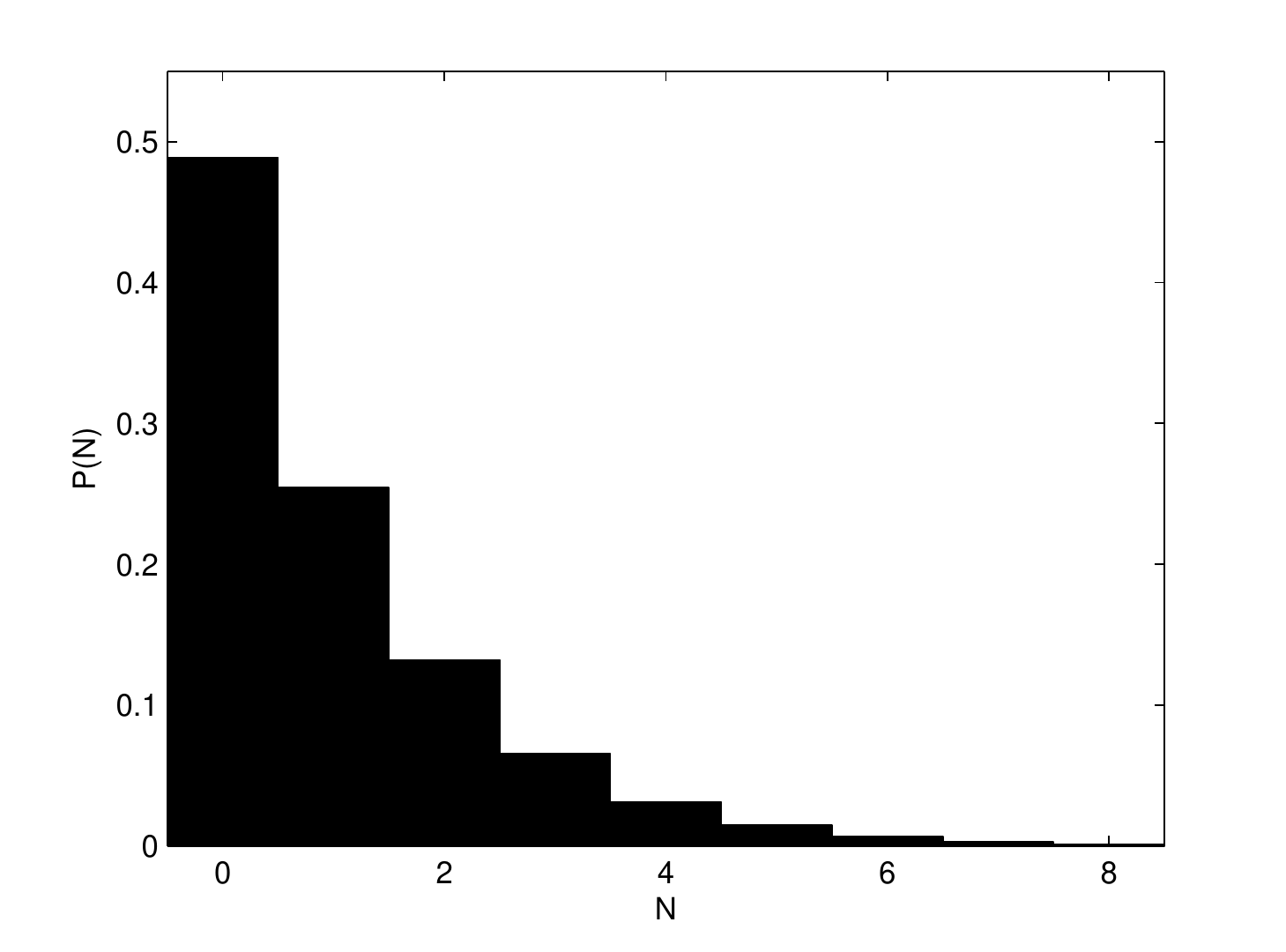}
\caption[Probability of $N$ observed bulge MSPs for lognormal luminosity distribution and X-shaped bulge model]{The probability distribution of observing $N$ MSPs from the bulge population based on the 
lognormal luminosity distribution and 
X-shaped bulge model.}
\label{fig:NObsLognormalX}
\end{figure}

\begin{figure} [!htb]
\centering
\subfigure{\centering\includegraphics[width=0.49\linewidth]{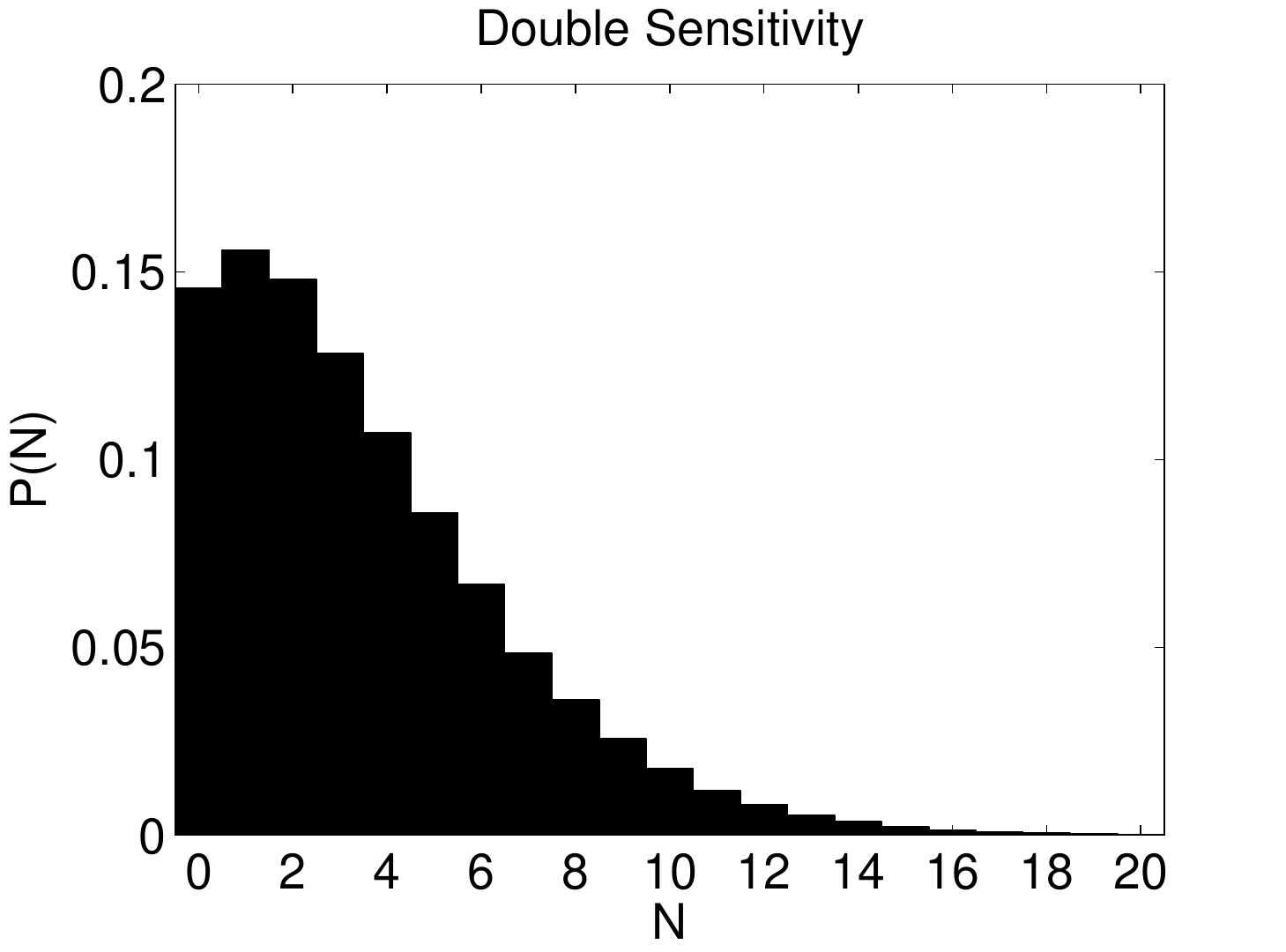}}
\subfigure{\centering\includegraphics[width=0.49\linewidth]{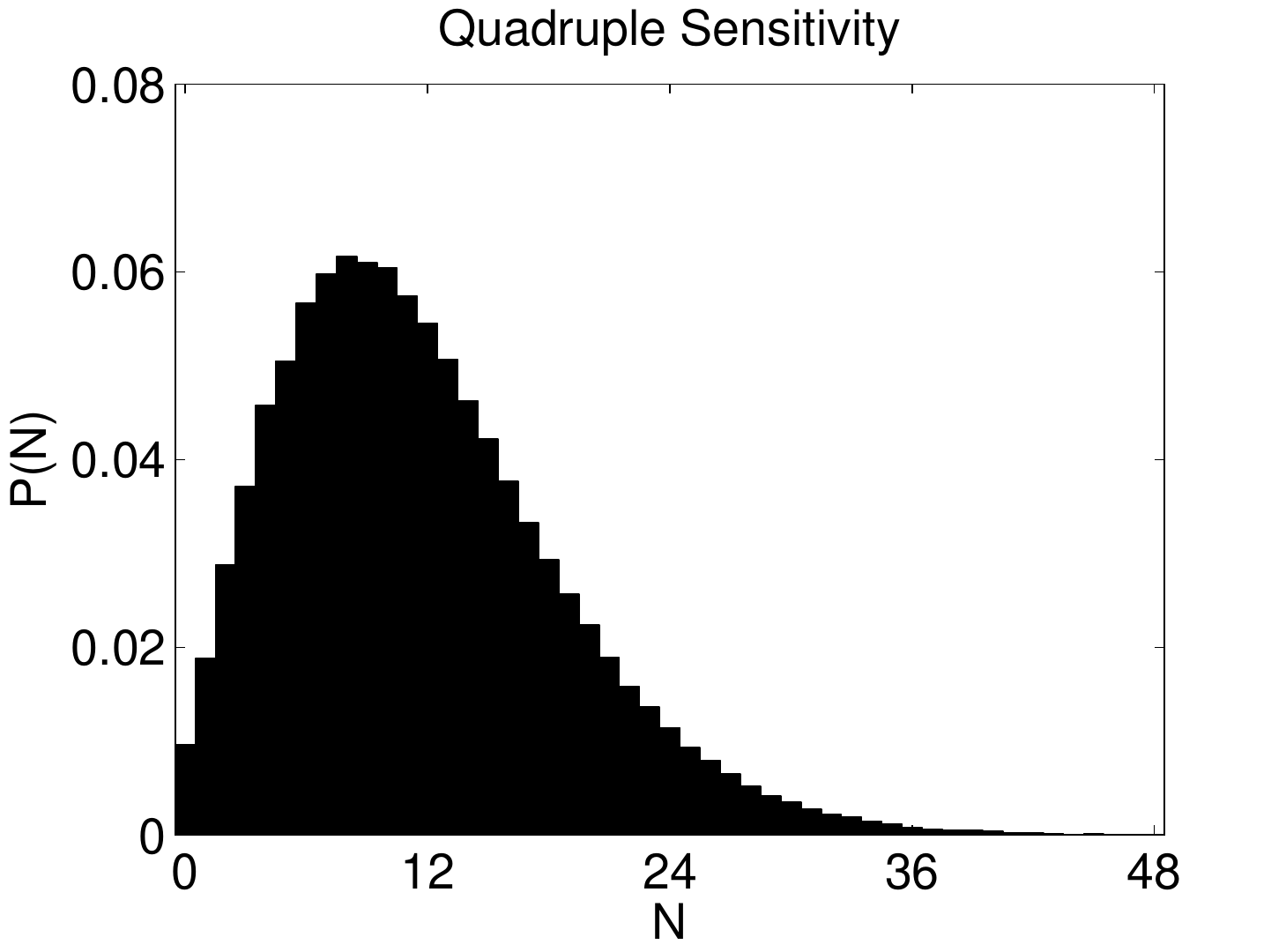}}
\caption[Probability of $N$ observed bulge MSPs for lognormal luminosity distribution and X-shaped bulge model with double or quadruple sensitivity]{The probability distribution of observing $N$ MSPs from the bulge population based on the lognormal luminosity distribution and X-shaped bulge model with double or quadruple sensitivity.}
\label{fig:NObsDoubleQuadrupleLognormalX}
\end{figure}

\begin{figure} [!htb]
\centering
\subfigure{\centering\includegraphics[width=0.49\linewidth]{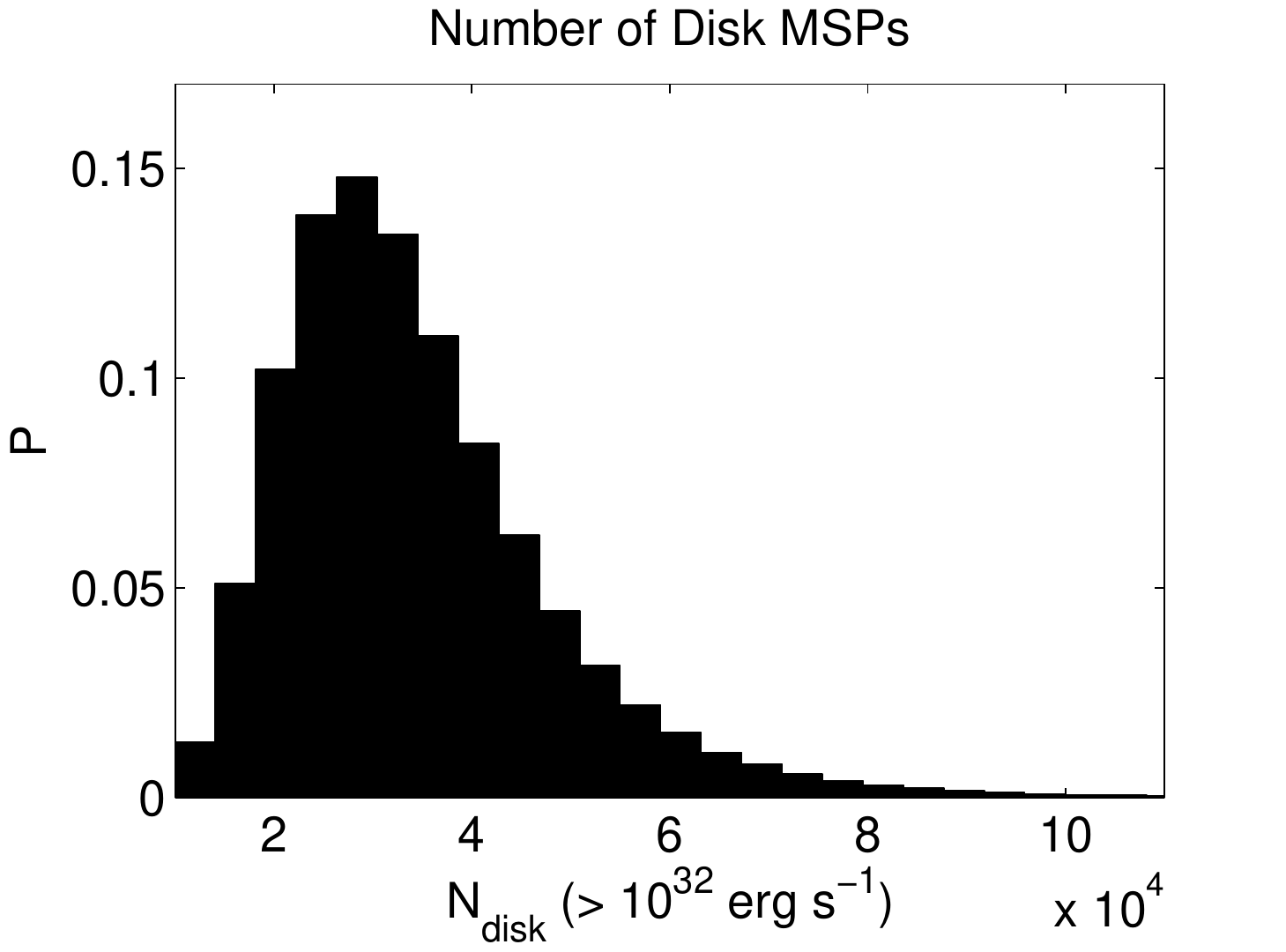}}
\subfigure{\centering\includegraphics[width=0.49\linewidth]{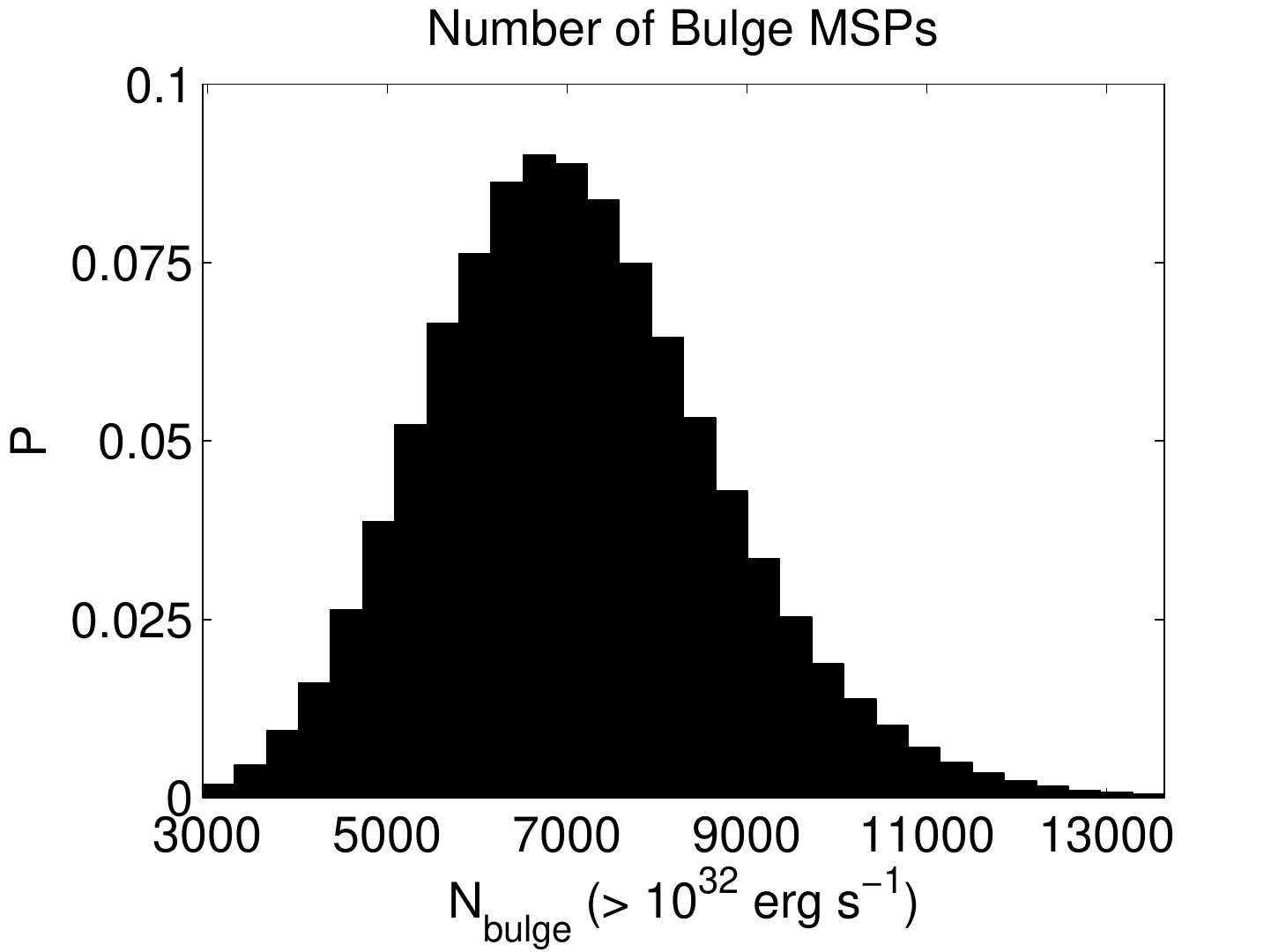}}
\caption[Number of MSPs for X-shaped bulge model and lognormal  luminosity function]{The distribution of the number of MSPs with luminosity greater than $10^{32}$ ${\rm erg \cdot s}^{-1}$ for the
lognormal  luminosity function and 
X-shaped bulge model.}
\label{fig:NumberMSPsLognormalX}
\end{figure}

The probabilities of observing one or more bulge MSPs and the expected number of observations are listed in table \ref{tab:ObsBulgeMSPLognormalX} with the probability distribution of observing $N$ bulge MSPs displayed in figures~\ref{fig:NObsLognormalX} and \ref{fig:NObsDoubleQuadrupleLognormalX}. The number of MSPs in the disk with luminosity greater than $10^{32}$ ${\rm erg \cdot s}^{-1}$ was $\left(3 \pm 1\right) \times 10^4$ and the number in the bulge was $\left(7 \pm 2\right) \times 10^3$. In figure~\ref{fig:NumberMSPsLognormalX}, the distribution of the number of MSPs simulated in the disk and bulge is shown.

\begin{table} [!htb]
\centering
\begin{tabular}{|c|c|c|}
\hline
Sensitivity Factor & Mean $N$ & P($N > 0$) \\ \hline \hline
1.0 & $0.999$ & $0.511$ \\ \hline
2.0 & $3.58$ & $0.854$ \\ \hline
4.0 & $11.7$ & $0.990$ \\ \hline
\end{tabular}

\caption[Probability of observing bulge MSPs and expected number for lognormal luminosity distribution and X-shaped bulge model]{Expected number of observed MSPs located in the bulge and probability of observing one or more for lognormal luminosity distribution and X-shaped bulge model.}
\label{tab:ObsBulgeMSPLognormalX}
\end{table}

\FloatBarrier

\section{Discussion}
\label{discussion}

Figures~\ref{fig:SimDataLognormalSpherical} and \ref{fig:SimDataLognormalX} 
show  simulated data produced using parameter sets in the Markov chains for the spherical and X-bulge. As can be seen in the top panels, the simulated distributions of resolved MSPs in longitude, latitude and flux are consistent with those observed. 
All the posterior predicted p-values were in the 0.025 to 0.975 required range corresponding to the $95\%$ error bar, except the 11th point of the last panel of figure~\ref{fig:SimDataLognormalX}, which has a predictive p-value of 0.021. The 12th point is very close with a p-value of 0.028.
However, these badly fitted points have minimal effect on the parameter fit values
as they are high energy GCE points and the parameters are mainly influenced by the lower energy GCE points which have a better signal to noise ratio. The two points may be indications of secondary emission from the bulge MSPS~\cite{Ioka}.

Although only distances measured using parallax were included as priors on the distance parameters, as can be seen from the left middle panel, the distance distributions of simulated resolved MSPs for both models are similar to that observed, including those distances estimated using the dispersion measure. 
This indicates the dispersion measure distance may not be biased.
The simulated parallax distance distributions are also good fits to the data. 

For the spherically symmetric bulge model, the overall goodness of fit p-values, calculated as described at the end of Section~\ref{methods}, were $0.7 \substack{+0.2 \\ -0.3}$ for resolved MSPs and $0.14 \substack{+0.13 \\ -0.02}$ for the GCE spectrum. For the X-shaped bulge model, these were $0.7 \substack{+0.2 \\ -0.3}$ and $0.57 \substack{+0.10 \\ -0.05}$ respectively. These results indicate that the model is an acceptable fit to the observed data.

The simulated flux distribution, seen in the middle right hand side panels, requires the uncertain flux threshold to fit the data. When instead of $F_{\rm th}$ being drawn from a lognormal distribution, $\sigma_{\rm th}$ was removed as a parameter and the threshold was simply $\exp(\mu_{\rm th}(l, b) + K_{\rm th})$, the MCMC simulation produces Markov chains with parameters which predict a significantly larger number of resolved low flux MSPs and few higher flux MSPs, resulting in a poor fit to the flux distribution. It is reassuring that our estimated number of disk MSPs in figure~\ref{fig:NumberMSPsLognormalSpherical} and  \ref{fig:NumberMSPsLognormalX} are consistent with the number of disk MSPs from radio observations \cite{Levin2013}. But a more detailed investigation is needed to properly compare the radio and gamma-ray results.

Comparing tables \ref{tab:ResultsLognormalSpherical} and \ref{tab:ResultsLognormalX} it can be seen that a mildly significant difference between the spherically symmetric and X-shaped bulge models is the ratio $r_{\rm d/b}$. The number of MSPs in the spherically symmetric bulge is similar to the number of disk MSPs. This is not the case when the X-shaped bulge model is used; here the number of disk MSPs is on average a factor of four larger. It can be seen that this difference is significant at the 3.8$\sigma$ level when accounting for the uncertainties in the parameter $\ln(r_{\rm d/b})$ by considering $\abs{\mu_1 - \mu_2}/\sqrt{\sigma_1^2 + \sigma_2^2}=3.8$ where $\mu_1$ and $\mu_2$ are the means with errors $\sigma_1$ and $\sigma_2$ for each of the bulge models. 

As can be seen from figures~\ref{fig:CornerLognormalSpherical} and \ref{fig:CornerLognormalX} there are correlations  between $\ln(r_{\rm d/b})$, $K_{\rm th}$ and $\sigma_{\rm th}$. There are likely two causes for this. The first is that if $K_{\rm th}$ increases, so must $\sigma_{\rm th}$; if it does not, the probability of observing an MSP with flux $F_i$, $p(F_i > F_{\rm th, i})$, may drop significantly for those MSPs where the measured flux is around the central flux threshold $\exp(\mu_{\rm th}(l_i, b_i) + K_{\rm th})$. The other cause of the correlation between the three variables is that if the flux threshold generally increases due to an increase in the parameter $K_{\rm th}$, the number of MSPs must increase to compensate for the decrease in resolved MSPs. Combined with the fact that bulge MSPs are unlikely to be observed -- which means $N_{\rm bulge}$ is largely dependent on the GCE and the luminosity function -- the increase occurs in $N_{\rm disk}$, causing the parameter $r_{\rm d/b}$ to be higher. The correlation between $\ln(r_{\rm d/b})$ and $\sigma_{\rm th}$ is then caused by the other two relationships.

There are also correlations between $\log_{10}(L_{\rm med})$ and $\sigma_L$. This occurs because the observed data could be explained either by a large underlying population of MSPs with a broad distribution of luminosities and a median lower than those observed or, alternatively, a smaller population of MSPs with a narrow luminosity distribution and median luminosity similar to the observed MSPs. It is likely the luminosity distribution is also, to some extent, constrained by the distribution of resolved MSPs in the sky. For example, a broad luminosity function with a relatively high median would tend to produce distant resolved MSPs that would be clustered in the direction of the GC. This occurs not only because of the bulge model, but also because the peak density of the disk spatial model is there. On the other hand, a narrow luminosity function with a relatively low median would result in the distribution of resolved MSPs being more evenly distributed in the sky (except at low latitudes due to the higher flux threshold). This is a result of the fact that MSPs which pass the flux threshold test would tend to be nearby, and therefore would be found in a volume throughout which the density of MSPs is approximately constant.

The spatial distribution is not particularly well constrained by the data, at the upper limit placed on $\sigma_r$ the likelihood remains relatively high. However, other work based on simulations and radio data suggests the radial distribution would not be as broad as that at this boundary \cite{Story:2007xy,Lorimer2013}.

The radial extent of the spherically symmetric bulge ($r_{\rm bulge}$) is determined from the  COBE-DIRBE NIR maps  to be around 3 kpc \cite{Mezger1996}.
Also, to make our results more easily comparable with ref.~\cite{HooperMohlabeng2016} we used $r_{\rm bulge}=3.1$~kpc. 
We  checked the sensitivity of the results to the radial extent of the spherically symmetric bulge by constructing a Markov chain for each of the cases where the bulge was larger or smaller in radius by 1 kpc. The only clear change in the parameter distribution was small shifts in $\ln(r_{\rm d/b})$, this likely occurred as when the bulge is spread out, the bulge MSP density will drop and so decrease the GCE unless the number of bulge MSPs is increased.
 The other change was in the expected number of resolved bulge MSPs, rising to 2.0 for a 4.1 kpc bulge radius and decreasing to 1.0 for the 2.1 kpc radius. As seen in Table \ref{tab:ObsBulgeMSPLognormalSpherical}, the expectation value was 1.4 for the 3.1 kpc bulge case.

\begin{figure}[!htb]
\includegraphics[width=1.0\linewidth]{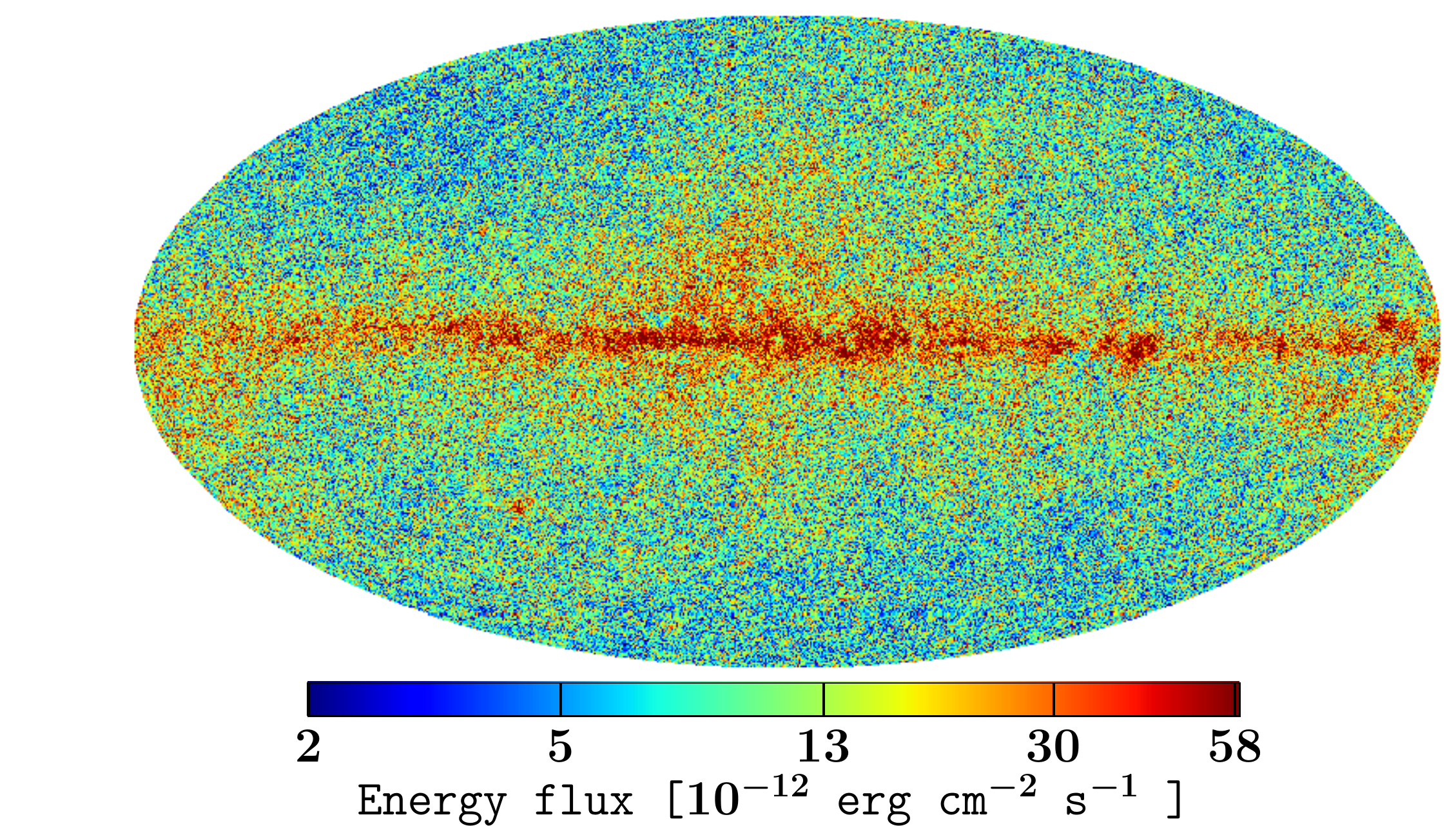}
\caption{\label{fig:StochasticThreshold}
Hammer-Aitoff projection of a sample threshold map, for flux above 100 MeV,  from eq.~(\ref{eq:flux_threshold_pdf}) using our mean posterior fit values from table~\ref{tab:ResultsLognormalSpherical}. For display purposes we have clipped the upper and lower 2.5\% pixels. 
}
\end{figure}

Our parameter fits for $K_{\rm th}$ and $\sigma_{\rm th}$ can be used to visualize the stochastic selection function we are using based on eq.~(\ref{eq:flux_threshold_pdf}). We illustrate a sample from this distribution in figure~\ref{fig:StochasticThreshold}.  Note that this image is only illustrative as if a pixel contains more than one MSP then each MSP in that pixel will be a separate draw from the distribution. This figure may be qualitatively compared with figure 16 of ref.~\cite{FermiPulsarsCatalog} which has also had some clipping applied. Although, note that each step in the Markov chain will generate a new sample of $K_{\rm th}$ and $\sigma_{\rm th}$ and so a new threshold map.

\begin{figure} [!htb]
\centering
\includegraphics[width=1.0\linewidth]{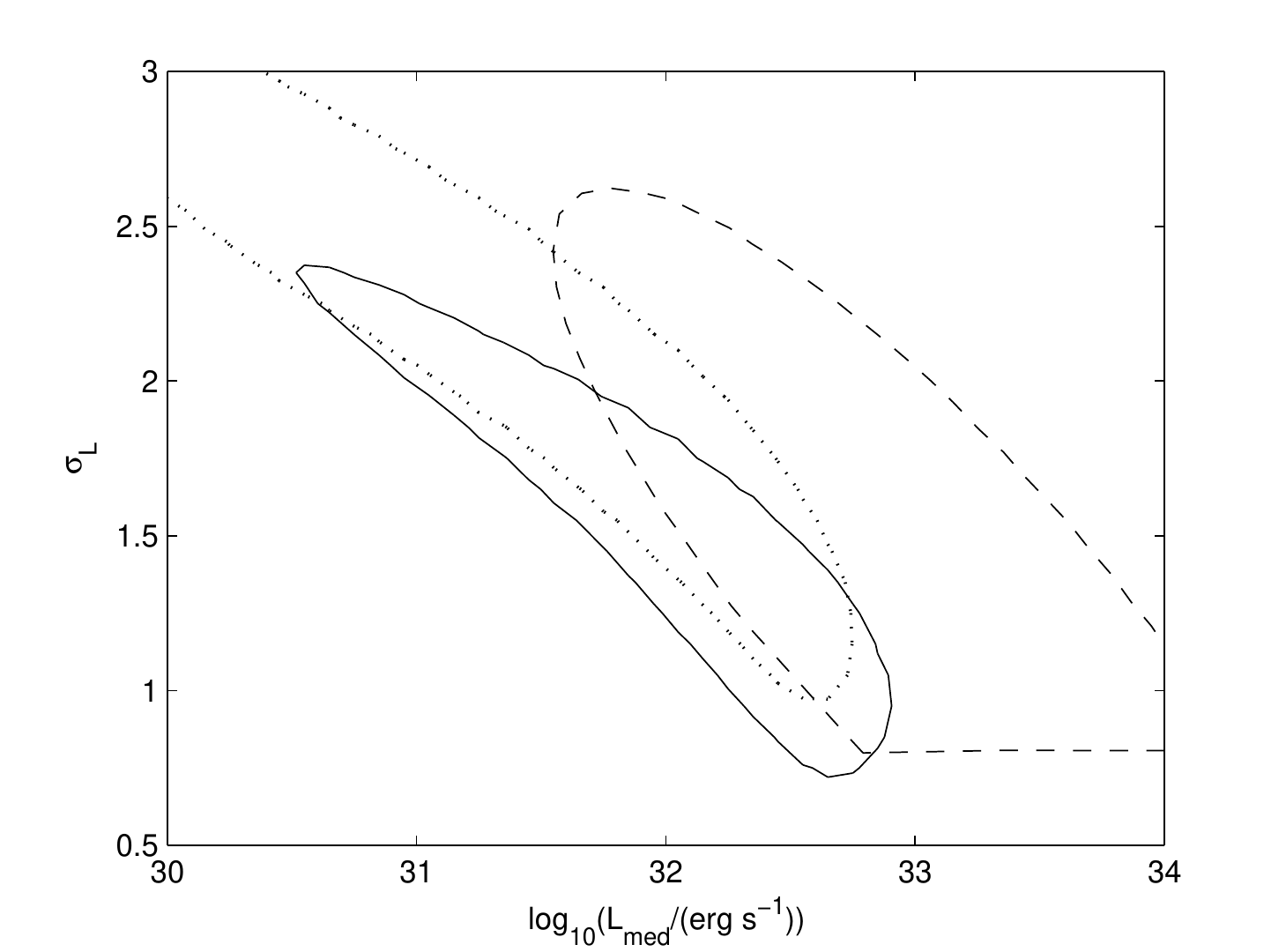}
\caption[Contour plot of luminosity function parameters for lognormal luminosity distribution and spherical bulge model]{Contour plot of luminosity function parameter distribution showing the $95\%$ contour for the lognormal luminosity distribution and 
spherical bulge model. The dashed line is the $2\sigma$ contour from ref.~\cite{HooperMohlabeng2016}. The dotted line is the $95\%$ contour for the luminosity function distribution assuming the spherical bulge does not contribute to resolved MSPs. }
\label{fig:LuminosityContourVsHooperLognormalSpherical}
\end{figure}

\begin{figure} [!htb]
\centering
\includegraphics[width=1.0\linewidth]{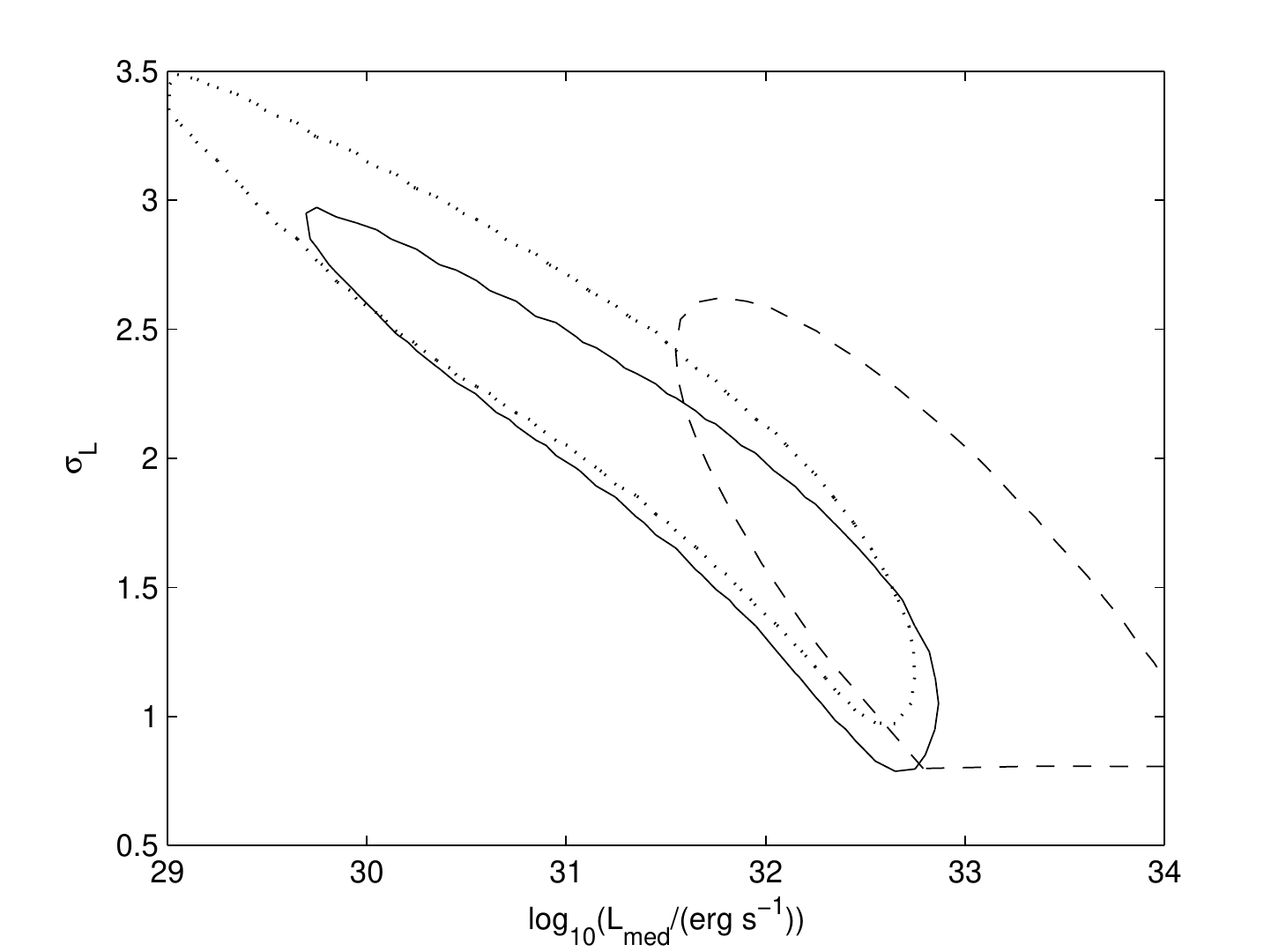}
\caption[Contour plot of luminosity function parameters for lognormal luminosity distribution and X-shaped bulge model]{Contour plot of luminosity function parameter distribution showing the $95\%$ contour for the lognormal luminosity distribution and X-shaped bulge model. The dashed line is the $2\sigma$ contour from ref.~\cite{HooperMohlabeng2016}. The dotted line is the $95\%$ contour for the luminosity function distribution assuming the spherical bulge does not contribute to resolved MSPs.}
\label{fig:LuminosityContourVsHooperLognormalX}
\end{figure}

\begin{figure} [!htb]
\centering
\includegraphics[width=1.0\linewidth]{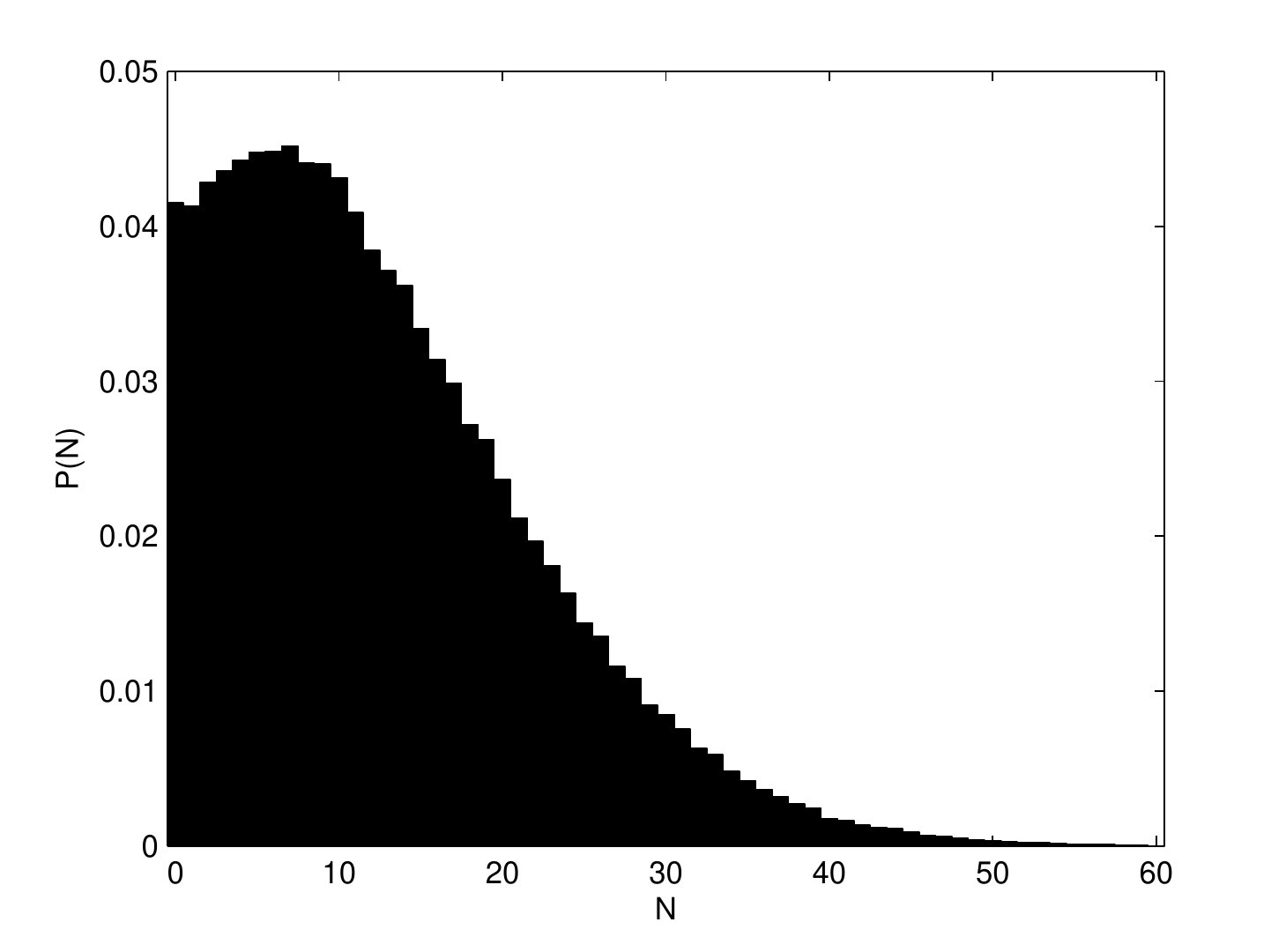}
\caption[Probability of observing $N$ bulge MSPs using luminosity function fitted for disk only]{The probability distribution of observing $N$ MSPs from the spherical bulge population based on a lognormal
luminosity function fitted assuming only MSPs from the disk model can be observed.}
\label{fig:NObsSphericalDiskOnly}
\end{figure}

Ref.~\cite{HooperMohlabeng2016} used the observed MSPs to attempt to find parameters for the lognormal luminosity function by using the product of three binned likelihoods one for longitude, one for latitude, and the other for flux \cite{Mohlabeng2016}. In each bin the expected number of observations $\lambda_i$ was found by taking a large number of random samples from the model, binning them, and normalizing so that $\sum_{i = 1}^{N} \lambda_i = 66$, where $66$ was the number of MSPs used in their fit. Their likelihood for each distribution was:

\begin{equation}
\mathcal{L} = \prod_{i} \frac{\lambda_i^{n_i} \exp(-\lambda_i)}{n_i!}
\end{equation}
where $n_i$ was the number of observed MSPs in bin $i$. The results are compared to the results of this work for both the spherically symmetric and X-shaped bulge models in figures~\ref{fig:LuminosityContourVsHooperLognormalSpherical} and \ref{fig:LuminosityContourVsHooperLognormalX}. While there is some agreement, the difference could be explained in part by the use of the parallax measured distances which were only used to estimate a lower bound on $\sigma_L$ in ref.~\cite{HooperMohlabeng2016}. The other significant difference between that study and this work is the separable form of likelihood they used. As seen from eq.~(\ref{eq:likelihood_unbinned}), the actual likelihood will not be separable in this way. We included, in addition,  the bulge model as part of the likelihood, not only to find parameters fitting the GCE, but also because some luminosity distributions could result in many observed bulge MSPs. This could reduce the likelihood for luminosity functions that tend to generate high luminosity MSPs with greater probability. Figures~\ref{fig:LuminosityContourVsHooperLognormalSpherical} and \ref{fig:LuminosityContourVsHooperLognormalX} also show the luminosity function parameter distribution which results when the contribution of resolved bulge MSPs is removed. Based on the fitted flux threshold, we found a mean of $12.8$ resolved spherical bulge MSPs, and the probability of observing one or more was $0.959$. The probability distribution is shown in figure~\ref{fig:NObsSphericalDiskOnly}. It is also suggested in ref.~\cite{HooperMohlabeng2016} that a further restriction can be placed on the luminosity function parameters by considering MSPs located in globular clusters. These were not taken into account here as it is unlikely  they would have the same luminosity distribution as the disk population.

It is concluded in ref.~\cite{HooperMohlabeng2016} that many MSPs located in the bulge should have already been observed if the luminosity distribution were the same for disk and bulge MSPs. While the probability of observing bulge MSPs varies depending on the model used as can be seen from tables~\ref{tab:ObsBulgeMSPLognormalSpherical} and \ref{tab:ObsBulgeMSPLognormalX},
we find that there is a significant probability that no bulge MSPs would have been resolved based on the fitted threshold and luminosity function parameters. Although six of the observed MSPs are inside the solid angle of the projected spherically symmetric bulge, the distances estimated using dispersion measures indeed indicate that none are actually located inside this bulge. No MSPs were observed in the region of the projected X-shaped bulge. With four times the current sensitivity to point sources it is likely that many bulge MSPs could be resolved. Double the current sensitivity could result in a few observations.

\begin{figure} [!htb]
\centering
\includegraphics[width=\linewidth]{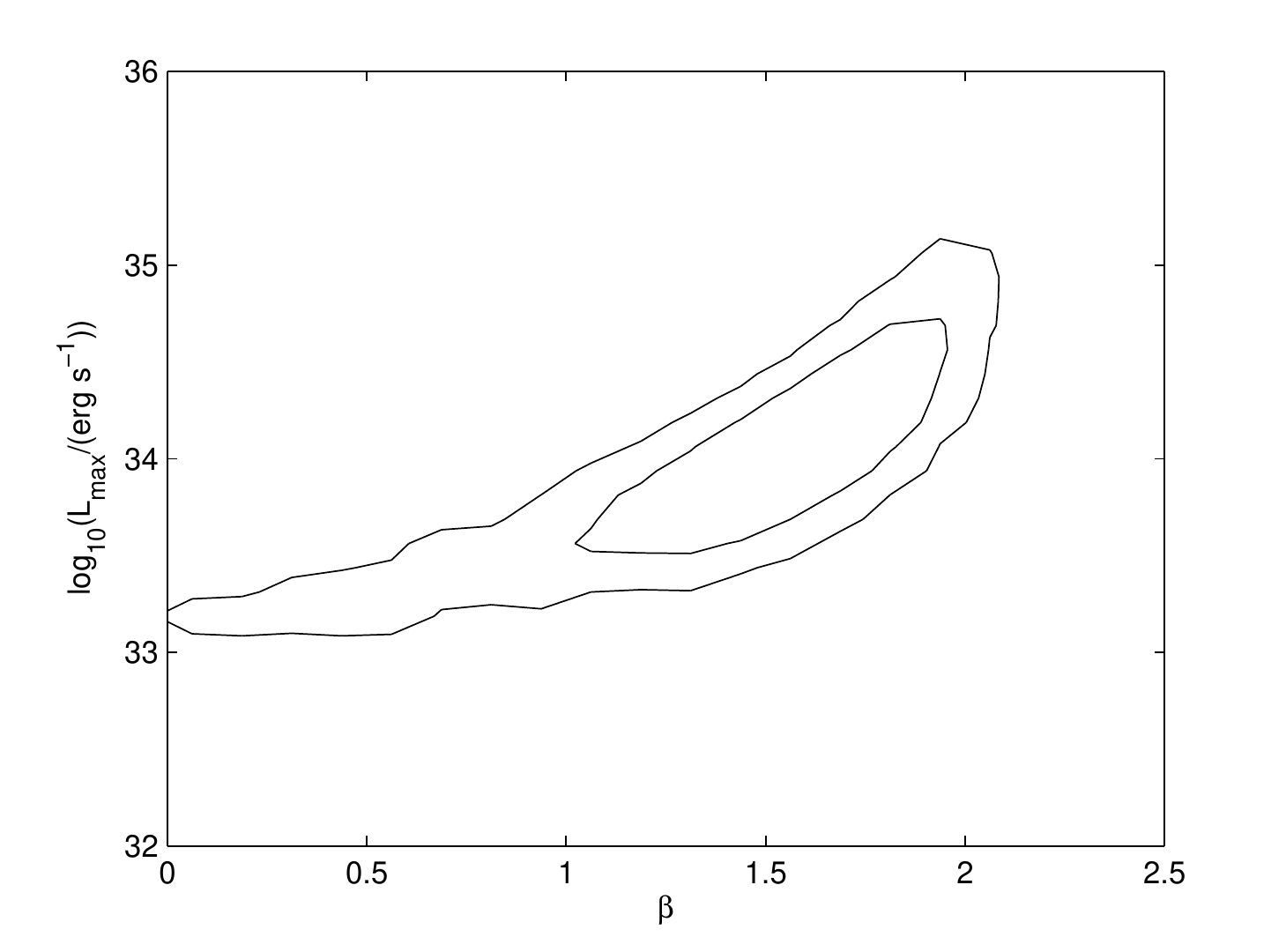}
\caption{Marginalized confidence intervals (68\% and 95\%) of a power law luminosity function, $dN/dL\propto L^{-\beta}$,
 with a maximum luminosity, $L_{\rm max}$. 
These constraints are for a disk only model of the resolved MSPs.
\label{fig:PL}
}
\end{figure}

As the work for this article was completed, a new article on a pulsar explanation of the GCE by the Fermi team was made available~\cite{Fermi-LAT:2017yoi}. In this new work, a power law was used to model the luminosity distribution of both young pulsars and MSPs combined. This was of the form $dN/dL \propto L^{-\beta}$ for $L$ between $L_{\rm min}=10^{31}$ ${\rm erg \cdot s}^{-1}$ and $L_{\rm max}=10^{36}$ ${\rm erg \cdot s}^{-1}$ and zero outside that range.  They found the luminosity distribution of nearby pulsars can be modeled by $\beta= 1.2$. 
To determine the number of disk and bulge pulsars, they took as pulsar candidates all detected point sources in a 40$^\circ$ by  40$^\circ$ area around the GC which had significantly curved spectra and when fitted, with an exponential cutoff spectrum, had a spectral slope less than 2 and an energy cutoff less than 10 GeV.  
To compare our resolved MSP constraints to their model,
 we fitted a disk only power law model with  the poorly constrained $L_{\rm min}$ fixed to $10^{31}$ erg/s and $L_{\rm max}$ and $\beta$ left as free parameters. 
Marginalizing over the other free parameters, in the same way as we did for the lognormal case, we got the constraints in figure~\ref{fig:PL}. Using the GCE of ref.~\cite{Dmitri2017}, we found the selection function of figure~\ref{fig:PL} had a mean number of resolved bulge MSPs which was $0.354$ and the probability of resolving one or more was $0.177$. But this does not imply the number of pulsar candidates based on the point source spectrum will be this low as ref.~\cite{Fermi-LAT:2017yoi} did not require the gamma-ray pulsations to be detected. As can be seen, for the resolved MSPs, a lower value of $L_{\rm max}$ than  ref.~\cite{Fermi-LAT:2017yoi} used is needed for $\beta=1.2$. Although, as they combined the MSPs and young pulsars, our result is not straight forwardly applicable to their case.


\section{Conclusions}
\label{conclusions}

An  MCMC algorithm was used to constrain parameters for a set of models of the Milky Way MSP population, with the luminosity distribution being of particular interest. 
This search over the parameter space was performed using the GCE data, the locations of observed MSPs in the sky, their fluxes, and their distances where parallax distance measurements were available.

To confirm that the models could plausibly explain the observations, the Markov chains of parameters produced by the MCMC simulation were then used to produce simulated data. The simulated distributions of resolved MSPs were a good fit to the observed data, as were the simulations of the GCE.

Although it seems unlikely that the disk and bulge MSPs would have the same luminosity function, our results show that the current data are not precise enough to reveal any differences between the two (if they exist). 
This is mainly because the bulge MSP population is relatively unconstrained by our analysis which is only sensitive to the luminosity of the entire population  via the product $\bar{L} N_{\rm bulge}$ where $\bar{L}$ is the average luminosity.
While the constraint that not too many bright, bulge MSPs be predicted does require that the luminosity function not be too skewed too much  towards high luminosities, this turns out not to be
a very tight constraint given the current point source sensitivity of observations towards the inner Galaxy.
As seen from tables~\ref{tab:ObsBulgeMSPLognormalSpherical} and \ref{tab:ObsBulgeMSPLognormalX}  an improvement in the gamma-ray point source detection sensitivity by a factor of around two may allow a number of MSPs from the bulge population to be resolved and hence help test the MSP proposal for the GCE. 


It was claimed by ref.~\cite{HooperMohlabeng2016} that if the GCE were produced by MSPs, many of them would already have been resolved. Here, it has been shown that this is in fact not the case. The main cause of this disagreement is likely due to ref.~\cite{HooperMohlabeng2016} not including the possibility of resolved bulge MSPs when they 
evaluated their best fit luminosity function parameters.


As the current article was being completed,
 ref.~\cite{Fermi-LAT:2017yoi} 
 presented an analysis based on
  new bulge pulsar candidates. But, these candidates have not been resolved in the sense of the current 3FGL resolved MSPs in that they do not yet have evidence of gamma-ray pulsations
  and have not been shown to be coincident with a radio pulsar.
In future work we plan to  include the  3FGL resolved young pulsars and the new pulsar candidates found in ref.~\cite{Fermi-LAT:2017yoi}  which will need  separate selection functions in addition to the one we are currently using in eq.~(\ref{eq:flux_threshold_pdf}). 
We will then use the new luminosity function parameters to update forecasts for  future observations, at radio wavelengths, of the bulge pulsars~\cite{Calore2016}.



\acknowledgments
We thank David Smith for providing the numerical values for figure~16 of ref.~\cite{FermiPulsarsCatalog} and also for helpful correspondence.
We gratefully acknowledge useful correspondence and conversation with Lilia Ferrario, Dan Hooper, Ivo Seitenzahl, Ashley Ruiter, Simone Scaringi, and Qiang Yuan.







\appendix

\section{Supplementary Material}
\label{appendix}
\subsection{Computational Methods}
To sample a random point from the disk spatial model of eq.~(\ref{eq:rho_disk}), $\abs{z}$ is drawn from an exponential distribution with scale parameter $z_0$, then $z$ is assigned either $\abs{z}$ or $-\abs{z}$ each with probability $0.5$. A value for $\theta$ is then drawn from a uniform distribution on the interval $\left[0, 2\pi\right)$. Finally, $r_{\rm cyl}$ is randomly drawn from the following probability distribution function:
\begin{equation}
\label{eq:disk_r_pdf}
f(r) = \frac{1}{\sigma_r^2} \exp(-r_{\rm cyl}^2/2\sigma_r^2) r_{\rm cyl}
\end{equation}
\noindent The factor of $r_{\rm cyl}$ is related to the volume element $\dd{V} = r_{\rm cyl} \dd{r_{\rm cyl}} \dd{z} \dd{\theta}$. 

Before the posterior distribution can be found it is necessary to find the number of MSPs in the disk and bulge implied by the parameters. Let $p_{\rm disk}$ and $p_{\rm bulge}$ be the probability of an MSP being observed in the disk and bulge respectively, then $N_{\rm disk}$  of eq.~(\ref{eq:rho_bulge})  and $N_{\rm bulge}$ of eq.~(\ref{eq:rho_disk}) can be found using the following two equations:

\begin{equation}
\label{eq:N_simultaneous_eqn}
\begin{split}
\lambda_{\rm res} &= p_{\rm disk} N_{\rm disk} + p_{\rm bulge} N_{\rm bulge} \\
r_{\rm d/b} &= N_{\rm disk} / N_{\rm bulge}
\end{split}
\end{equation}
\noindent Solving for $N_{\rm disk}$ and $N_{\rm bulge}$ gives:

\begin{equation}
\label{eq:N_disk_N_bulge}
\begin{split}
N_{\rm disk} &= \frac{\lambda_{\rm res}}{p_{\rm disk} + p_{\rm bulge} / r_{\rm d/b}} \\
N_{\rm bulge} &= \frac{\lambda_{\rm res}}{p_{\rm disk} r_{\rm d/b} + p_{\rm bulge}}
\end{split}
\end{equation}

\noindent It is not practical to find $p_{\rm disk}$ and $p_{\rm bulge}$ exactly as this would require a multidimensional integral for every iteration of the  MCMC algorithm. A simplistic way to approximate $p_{\rm disk}$ or $p_{\rm bulge}$ would be to draw a large number of simulated MSPs from the corresponding model (disk or bulge) and use the fraction that are resolved. However, this method has a significant disadvantage: the number of MSPs that it would be necessary to simulate to ensure the result is reasonably accurate could potentially be extremely large. This is because the number of simulated MSPs that are resolved is Poisson distributed, therefore for a relative error of on the order of $1\%$ we may wish to continue drawing from the disk model until we have several thousand resolved (usually $p_{\rm bulge} \ll p_{\rm disk}$ and $r_{\rm d/b} \sim 1$ so accuracy is less important for the bulge), but for luminosity functions which produce few highly luminous MSPs this could mean millions of draws from the model.

There is a simple improvement that can be made by recognizing that position in $l$, $b$ and $d$ allows a luminosity threshold $L_{\rm th}$ to be found. Before running the  MCMC algorithm for each Markov chain, distributions of points were generated according to the  bulge  and disk spatial models and the position of each point was converted to $l$, $b$ and $d$, in addition, for each point, a random number $u$ was drawn from the unit normal distribution to allow us to account for the uncertain flux threshold. Using eq.~(\ref{eq:flux_luminosity}) and the fact that the logarithm of the flux threshold is normally distributed with scale parameter $\sigma_{\rm th}$, the luminosity threshold for point $i$ is:

\begin{equation}
\label{eq:lum_thresh}
L_{{\rm th,}i} =4 \pi d_i^2 \exp(\mu_{\rm th}(l_i, b_i) + K_{\rm th} + \sigma_{\rm th} u_i)
\end{equation}

\noindent For any proposed set of parameters we can use importance sampling \cite{NumericalRecipes} to evaluate  the probability of a single random MSP being observed:

\begin{equation}
\label{eq:prob_resolved}
p \approx \frac{1}{N} \sum_{i = 1}^{N} w_i p(L > L_{{\rm th,}i})
\end{equation}

\noindent where $N$ is the total number of points, $p(L > L_{{\rm th,}i})$ is the probability that a randomly generated luminosity is greater than the threshold and $p$ represents either $p_{\rm disk}$ or $p_{\rm bulge}$. As the spatial parameters are allowed to vary, the spatial distribution of MSPs we are interested in is not the same as that which the points were sampled from. For point $i$, $w_i$ is the spatial density of MSPs given by the parameters divided by the density of points. Because the same spatial distribution of points and distribution of $u_i$ is used for each iteration of the algorithm, it can be guaranteed that a particular set of parameters will always give the same result for $N_{\rm disk}$ and $N_{\rm bulge}$. A further improvement that was made involved generating nearer points with higher probability. This ensures a larger proportion of points have relatively low $L_{{\rm th,}i}$. Without this, if the luminosity function parameters give a distribution heavily weighted towards lower luminosities, $p$ might effectively depend on a relatively small number of points. Not only could this result in larger errors, but if this issue is resolved by simply generating more points, it also means a large amount of time is spent evaluating $p(L > L_{{\rm th,}i})$ which have a negligible contribution.


\begin{landscape}
\subsection{Resolved MSP Data}

\begin{longtable}{|c|c|c|c|c|c|c|c|c|}
\hline
Name & $l$ & $b$ & Flux & $\Gamma$ (or $\omega$) & $E_{\rm cut}$ & $\beta$ & $E_0$ & $d$ \\
& (deg) & (deg) & ($10^{-11}$ erg cm$^{-2}$ s$^{-1}$) & & (GeV) & & (GeV) & (kpc) \\ \hline \hline
\endhead
J0023+0923 & $111.5$ & $-52.9$ & $0.73 \pm 0.08$ & $1.0 \pm 0.3$ & $1.0 \pm 0.2$ & - & $0.90$ & $0.7 \pm 0.1$ \\ \hline
J0030+0451 & $113.1$ & $-57.6$ & $6.1 \pm 0.2$ & $1.28 \pm 0.05$ & $2.1 \pm 0.2$ & - & $0.76$ & $0.28 \substack{+0.1 \\ -0.06}$ (P) \\ \hline
J0034-0534 & $111.5$ & $-68.1$ & $1.8 \pm 0.1$ & $1.7 \pm 0.1$ & $2.9 \pm 0.6$ & - & $0.70$ & $0.5 \pm 0.1$ \\ \hline
J0101-6422 & $301.2$ & $-52.7$ & $1.25 \pm 0.09$ & $1.5 \pm 0.1$ & $2.9 \pm 0.6$ & - & $0.90$ & $0.55 \substack{+0.09 \\ -0.08}$ \\ \hline
J0102+4839 & $124.9$ & $-14.2$ & $1.7 \pm 0.1$ & $1.8 \pm 0.1$ & $6 \pm 2$ & - & $1.2$ & $2.3 \substack{+0.5 \\ -0.4}$ \\ \hline
J0218+4232 & $139.5$ & $-17.5$ & $4.8 \pm 0.2$ & $1.94 \pm 0.06$ & $4.7 \pm 0.8$ & - & $0.74$ & $2.7 \substack{+1.1 \\ -0.7}$ \\ \hline
J0251+26 & $153.9$ & $-29.5$ & $0.7 \pm 0.1$ & $2.3 \pm 0.1$ & - & - & $1.1$ & $0.8 \substack{+0.2 \\ -0.1}$ \\ \hline
J0307+7443 & $131.7$ & $14.2$ & $1.46 \pm 0.08$ & $0.5$ & $1.24 \pm 0.07$ & - & $1.9$ & - \\ \hline
J0340+4130 & $153.8$ & $-11.0$ & $2.2 \pm 0.1$ & $1.3 \pm 0.1$ & $3.8 \pm 0.7$ & - & $1.4$ & $1.7 \pm 0.3$ \\ \hline
J0437-4715 & $253.4$ & $-42.0$ & $1.79 \pm 0.09$ & $1.3 \pm 0.1$ & $1.0 \pm 0.2$ & - & $0.59$ & $0.156 \pm 0.001$ (P) \\ \hline
J0533+6759 & $144.8$ & $18.2$ & $0.96 \pm 0.09$ & $1.6 \pm 0.2$ & $6 \pm 2$ & - & $1.4$ & $2.4 \substack{+0.9 \\ -0.6}$ \\ \hline
J0605+3757 & $174.2$ & $8.0$ & $0.7 \pm 0.1$ & $0.7 \pm 0.4$ & $1.5 \pm 0.4$ & - & $1.5$ & $0.7 \pm 0.1$ \\ \hline
J0610-2100 & $227.7$ & $-18.2$ & $1.1 \pm 0.1$ & $2.34 \pm 0.07$ & - & - & $0.85$ & - \\ \hline
J0613-0200 & $210.4$ & $-9.3$ & $3.4 \pm 0.2$ & $1.4 \pm 0.1$ & $2.9 \pm 0.4$ & - & $1.1$ & $0.9 \substack{+0.4 \\ -0.2}$ (P) \\ \hline
J0614-3329 & $240.5$ & $-21.8$ & $11.1 \pm 0.2$ & $1.36 \pm 0.03$ & $4.7 \pm 0.3$ & - & $0.88$ & $1.9 \pm 0.4$ \\ \hline
J0621+2514 & $187.1$ & $5.1$ & $1.1 \pm 0.1$ & $2.17 \pm 0.09$ & - & - & $1.7$ & $2.3 \substack{+0.5 \\ -0.4}$ \\ \hline
J0751+1807 & $202.8$ & $21.1$ & $1.3 \pm 0.1$ & $1.3 \pm 0.2$ & $3.0 \pm 0.6$ & - & $1.2$ & $0.4 \substack{+0.2 \\ -0.1}$ (P) \\ \hline
J0931-1902 & $251.0$ & $23.0$ & $0.30 \pm 0.09$ & $1.8 \pm 0.2$ & - & - & $3.4$ & $1.9 \substack{+0.5 \\ -0.4}$ \\ \hline
J0955-61 & $283.7$ & $-5.7$ & $0.8 \pm 0.1$ & $2.3 \pm 0.1$ & - & - & $1.2$ & $4.0 \substack{+0.9 \\ -0.8}$ \\ \hline
J1024-0719 & $251.7$ & $40.5$ & $0.36 \pm 0.05$ & $0.5$ & $1.4 \pm 0.2$ & - & $2.1$ & $0.39 \pm 0.04$ \\ \hline
J1124-3653 & $284.1$ & $22.8$ & $1.3 \pm 0.1$ & $1.3 \pm 0.2$ & $3.1 \pm 0.7$ & - & $1.4$ & $1.7 \pm 0.4$ \\ \hline
J1125-5825 & $291.8$ & $2.6$ & $1.5 \pm 0.3$ & $2.4 \pm 0.1$ & - & - & $1.9$ & $2.6 \pm 0.4$ \\ \hline
J1125-6014 & $292.5$ & $0.9$ & $1.2 \pm 0.3$ & $2.4 \pm 0.1$ & - & - & $2.1$ & $1.5 \pm 0.2$ \\ \hline
J1137+7528 & $129.1$ & $40.8$ & $0.23 \pm 0.06$ & $2.3 \pm 0.2$ & - & - & $1.5$ & - \\ \hline
J1142+0119 & $267.6$ & $59.4$ & $0.62 \pm 0.08$ & $1.2 \pm 0.3$ & $5 \pm 2$ & - & $1.8$ & $0.9 \pm 0.2$ \\ \hline
J1207-5050 & $295.9$ & $11.4$ & $0.8 \pm 0.1$ & $2.1 \pm 0.1$ & - & - & $1.5$ & $1.5 \pm 0.2$ \\ \hline
J1227-4853 & $299.0$ & $13.8$ & $4.1 \pm 0.2$ & $2.23 \pm 0.05$ & - & $0.11 \pm 0.03$ & $0.44$ & $1.4 \pm 0.2$  \\ \hline
J1231-1411 & $295.5$ & $48.4$ & $10.3 \pm 0.2$ & $1.18 \pm 0.04$ & $2.6 \pm 0.2$ & - & $0.92$ & $0.44 \pm 0.05$ \\ \hline
J1301+0833 & $310.8$ & $71.3$ & $1.1 \pm 0.1$ & $2.4 \pm 0.1$ & - & - & $0.77$ & $0.7 \pm 0.1$ \\ \hline
J1302-3258 & $305.6$ & $29.8$ & $1.1 \pm 0.1$ & $1.1 \pm 0.2$ & $2.7 \pm 0.7$ & - & $1.4$ & $1.0 \pm 0.2$ \\ \hline
J1311-3430 & $307.7$ & $28.2$ & $6.5 \pm 0.2$ & $1.91 \pm 0.05$ & $5.1 \pm 0.8$ & - & $0.66$ & $1.4 \pm 0.3$ \\ \hline
J1312+0051 & $314.8$ & $63.2$ & $1.6 \pm 0.1$ & $1.5 \pm 0.1$ & $2.8 \pm 0.6$ & - & $0.84$ & $0.8 \substack{+0.2 \\ -0.1}$ \\ \hline
J1400-1438 & $326.9$ & $45.0$ & $0.9 \pm 0.1$ & $2.30 \pm 0.08$ & - & - & $0.95$ & $0.48 \pm 0.03$ \\ \hline
J1446-4701 & $322.5$ & $11.4$ & $1.3 \pm 0.1$ & $2.30 \pm 0.08$ & - & - & $1.0$ & $1.5 \pm 0.2$ \\ \hline
J1514-4946 & $325.2$ & $6.8$ & $4.3 \pm 0.2$ & $1.43 \pm 0.09$ & $5.0 \pm 0.7$ & - & $1.5$ & $0.9 \pm 0.1$ \\ \hline
J1536-4948 & $328.2$ & $4.8$ & $8.7 \pm 0.3$ & $1.87 \pm 0.03$ & - & $0.17 \pm 0.02$ & $1.0$ & $1.8 \pm 0.1$ \\ \hline
J1543-5149 & $327.9$ & $2.7$ & $2.2 \pm 0.3$ & $2.5 \pm 0.1$ & - & - & $0.83$ & $2.4 \pm 0.2$ \\ \hline
J1544+4937 & $79.2$ & $50.2$ & $0.36 \pm 0.06$ & $2.3 \pm 0.1$ & - & - & $1.1$ & $1.2 \substack{+0.4 \\ -0.3}$ \\ \hline
J1600-3053 & $344.1$ & $16.5$ & $0.6 \pm 0.1$ & $1.0 \pm 0.5$ & $3 \pm 2$ & - & $2.6$ & $1.6 \pm 0.3$ \\ \hline
J1614-2230 & $352.6$ & $20.2$ & $2.3 \pm 0.1$ & $0.8 \pm 0.2$ & $1.8 \pm 0.3$ & - & $1.4$ & $0.65 \pm 0.05$ (P) \\ \hline
J1628-3205 & $347.4$ & $11.5$ & $1.2 \pm 0.1$ & $2.3 \pm 0.1$ & - & $0.4 \pm 0.1$ & $1.0$ & $1.3 \pm 0.2$ \\ \hline
J1630+3734 & $60.2$ & $43.3$ & $0.7 \pm 0.1$ & $1.0 \pm 0.4$ & $2.0 \pm 0.6$ & - & $1.6$ & $0.9 \pm 0.1$ \\ \hline
J1658-5324 & $334.9$ & $-6.6$ & $2.0 \pm 0.2$ & $1.4 \pm 0.2$ & $1.4 \pm 0.3$ & - & $1.0$ & $0.9 \pm 0.1$ \\ \hline
J1713+0747 & $28.8$ & $25.2$ & $0.9 \pm 0.1$ & $1.4 \pm 0.3$ & $3 \pm 1$ & - & $1.4$ & $1.05 \substack{+0.06 \\ -0.05}$ \\ \hline
J1732-5049 & $340.0$ & $-9.4$ & $0.9 \pm 0.1$ & $2.4 \pm 0.1$ & - & - & $0.96$ & $1.4 \pm 0.2$ \\ \hline
J1741+1351 & $37.9$ & $21.6$ & $0.6 \pm 0.1$ & $2.2 \pm 0.1$ & - & - & $1.3$ & $1.08 \substack{+0.04 \\ -0.05}$ (P) \\ \hline
J1744-1134 & $14.8$ & $9.2$ & $3.9 \pm 0.2$ & $1.4 \pm 0.1$ & $1.5 \pm 0.2$ & - & $0.75$ & $0.42 \pm 0.02$ (P) \\ \hline
J1745+1017 & $34.9$ & $19.3$ & $1.1 \pm 0.1$ & $1.5 \pm 0.3$ & $4 \pm 1$ & - & $1.5$ & $1.3 \substack{+0.2 \\ -0.1}$ \\ \hline
J1747-4036 & $350.2$ & $-6.4$ & $1.6 \pm 0.2$ & $2.3 \pm 0.1$ & - & - & $1.2$ & $3.4 \pm 0.8$ \\ \hline
J1805+06 & $33.4$ & $13.0$ & $0.5 \pm 0.1$ & $1.7 \pm 0.3$ & - & $0.6 \pm 0.2$ & $1.9$ & $2.5 \substack{+0.5 \\ -0.4}$ \\ \hline
J1810+1744 & $44.6$ & $16.8$ & $2.2 \pm 0.1$ & $1.8 \pm 0.1$ & $2.9 \pm 0.7$ & - & $0.66$ & $2.0 \substack{+0.3 \\ -0.2}$ \\ \hline
J1816+4510 & $72.9$ & $24.8$ & $1.21 \pm 0.09$ & $1.7 \pm 0.1$ & $5 \pm 1$ & - & $0.91$ & $2.4 \substack{+0.7 \\ -0.4}$ \\ \hline
J1843-1113 & $22.0$ & $-3.4$ & $2.0 \pm 0.3$ & $2.7 \pm 0.1$ & - & - & $0.50$ & $1.7 \pm 0.2$ \\ \hline
J1858-2216 & $13.6$ & $-11.4$ & $0.8 \pm 0.1$ & $0.7 \pm 0.4$ & $1.8 \pm 0.6$ & - & $1.6$ & $0.9 \pm 0.1$ \\ \hline
J1902-5105 & $345.6$ & $-22.4$ & $2.1 \pm 0.1$ & $1.8 \pm 0.1$ & $3.5 \pm 0.9$ & - & $0.62$ & $1.2 \pm 0.2$ \\ \hline
J1903-7051 & $324.4$ & $-26.5$ & $1.2 \pm 0.1$ & $2.32 \pm 0.07$ & - & - & $0.91$ & $0.8 \pm 0.1$ \\ \hline
J1959+2048 & $59.2$ & $-4.7$ & $1.8 \pm 0.2$ & $1.3 \pm 0.2$ & $1.4 \pm 0.3$ & - & $0.80$ & $2.5 \substack{+0.2 \\ -0.5}$ \\ \hline
J2017+0603 & $48.6$ & $-16.0$ & $3.5 \pm 0.2$ & $1.1 \pm 0.1$ & $3.7 \pm 0.5$ & - & $1.7$ & $1.6 \pm 0.2$ \\ \hline
J2042+0246 & $49.0$ & $-23.0$ & $0.36 \pm 0.06$ & $2.3 \pm 0.3$ & - & $1$ & $1.2$ & - \\ \hline
J2043+1711 & $61.9$ & $-15.3$ & $3.0 \pm 0.1$ & $1.58 \pm 0.08$ & $4.5 \pm 0.8$ & - & $0.96$ & $1.8 \substack{+0.1 \\ -0.3}$ \\ \hline
J2047+1053 & $57.1$ & $-19.6$ & $0.36 \pm 0.06$ & $0.5$ & $1.7 \pm 0.3$ & - & $2.6$ & $2.1 \pm 0.3$ \\ \hline
J2051-0827 & $39.2$ & $-30.5$ & $0.32 \pm 0.05$ & $0.5$ & $1.6 \pm 0.3$ & - & $2.4$ & $1.0 \substack{+0.2 \\ -0.1}$ \\ \hline
J2124-3358 & $10.9$ & $-45.4$ & $3.9 \pm 0.1$ & $0.9 \pm 0.1$ & $1.9 \pm 0.2$ & - & $1.1$ & $0.30 \substack{+0.07 \\ -0.05}$ (P) \\ \hline
J2129-0429 & $48.9$ & $-36.9$ & $1.0 \pm 0.1$ & $2.22 \pm 0.07$ & - & - & $1.0$ & $0.9 \pm 0.1$ \\ \hline
J2214+3000 & $86.9$ & $-21.7$ & $3.3 \pm 0.1$ & $1.20 \pm 0.08$ & $2.4 \pm 0.3$ & - & $0.94$ & $1.5 \pm 0.2$ \\ \hline
J2215+5135 & $99.9$ & $-4.2$ & $1.4 \pm 0.1$ & $1.4 \pm 0.2$ & $3.5 \pm 0.8$ & - & $1.3$ & $3.0 \substack{+0.3 \\ -0.4}$ \\ \hline
J2234+0944 & $76.3$ & $-40.4$ & $0.8 \pm 0.1$ & $1.3 \pm 0.3$ & $2.1 \pm 0.7$ & - & $1.1$ & - \\ \hline
J2241-5236 & $337.4$ & $-54.9$ & $3.1 \pm 0.1$ & $1.25 \pm 0.08$ & $2.8 \pm 0.4$ & - & $0.94$ & $0.51 \pm 0.08$ \\ \hline
J2256-1024 & $59.2$ & $-58.2$ & $0.77 \pm 0.08$ & $1.2 \pm 0.2$ & $2.3 \pm 0.6$ & - & $1.2$ & $0.6 \pm 0.1$ \\ \hline
J2302+4442 & $103.4$ & $-14.0$ & $3.8 \pm 0.1$ & $1.19 \pm 0.08$ & $3.0 \pm 0.3$ & - & $1.1$ & $1.2 \substack{+0.1 \\ -0.2}$ \\ \hline
J2339-0533 & $81.3$ & $-62.5$ & $3.0 \pm 0.1$ & $1.51 \pm 0.08$ & $5.1 \pm 0.9$ & - & $0.96$ & - \\ \hline

\caption[Observed millisecond pulsars]{List of observed millisecond pulsars. For the spectral parameters, MSPs with neither the $E_{\rm cut}$ or $\beta$ columns filled have been fitted with a simple power law spectrum, those with $E_{\rm cut}$ filled have been fitted with an exponentially cutoff power law, and those with the $\beta$ column filled have been fitted with a log parabolic spectrum. Data from 3FGL catalog \cite{Acero15} except distance measurements. A (P) in the distance column indicates the distance is that reported in ref.~\cite{FermiPulsarsCatalog} and was found using the parallax method. Other distances were derived from the dispersion measure reported in the ATNF pulsar catalog \cite{ATNF:catalog} using the NE2001 model \cite{Cordes02}. }
\label{tab:msp_data}
\end{longtable}
\end{landscape}




\bibliography{MSPs,thesis_bib}

\end{document}